\documentclass[12pt]{article}

\usepackage{epsfig}

\begin{document}

\baselineskip=21pt

\begin{Large}
\noindent {\bf A toy model of the five-dimensional universe 
      with the cosmological constant}
\end{Large}

\bigskip	
\bigskip

\baselineskip=0.2077in

\begin{large}
\noindent Wojciech Tarkowski 
\end{large}

\bigskip

\noindent {\it St.\ Paul Research Laboratory, ul.\ Konwaliowa 12,
05-123 Chotom\'ow, Poland
\\ e-mail:}\ {\tt tarkowski@data.pl}

\sloppy
\date{}

\bigskip
\bigskip

\noindent {\bf Abstract}

\medskip

\noindent {\small
A value of the cosmological constant in a toy model of the 
five-dimensional universe is calculated in such a manner that it 
remains in agreement with both astronomical observations and 
the quantum field theory concerning the zero-point fluctuations 
of the vacuum. The (negative) cosmological constant is equal
to the inverse of the Planck length squared, which means
that in the toy model the vanishing of the {\it observed}
value of the cosmological constant is a consequence of the existence
of an energy cutoff exactly at the level of the Planck scale. In turn,
a model for both a virtual and a real particle--antiparticle pair
is proposed which describes properly some energetic properties
of both the vacuum fluctuations and created particles, as well
as it allows one to calculate the {\it discrete} ``bare'' values of 
an elementary-particle mass, electric charge and intrinsic angular 
momentum (spin) at the energy cutoff. The relationships between the 
discussed model and some phenomena such as the {\it Zitterbewegung} 
and the Unruh--Davies effect are briefly analyzed, too. The proposed 
model also allows one to derive the Lorentz transformation and the
Maxwell equations while considering the properties of the vacuum
filled with the sea of virtual particles and their antiparticles.
Finally, the existence of a finite value of the 
vacuum-energy density resulting from the toy model leads us to the
formulation of dimensionless Einstein field equations which can be
derived from the Lagrangian with a dimensionless (na\"{\i}vely 
renormalized) coupling constant.}

\bigskip
\bigskip
\bigskip
\medskip

\baselineskip=0.18in


\smallskip


\vspace{0.5cm}

\baselineskip=0.225in

\newpage

\noindent {\bf 1.\ Introduction}
\bigskip

Various astronomical observations place strong limits on the absolute 
value of the cosmological constant $\lambda$ requiring it should not 
be greater than {\it c\`a.}\ $10^{-50}$ m$^{-2}$; see
Ref.~\cite{Peebles}.  On the other hand, the quantum theory
predicts that anything contributing to the vacuum-energy density
should act like a cosmological constant. Theoretical expectations thus 
give the value of $\lambda$ of the order of $\pm 10^{70}$ m$^{-2}$, 
which exceeds observational limits by about $120$ orders of magnitude.
This enormous discrepancy is at the origin of a dilemma often
referred to as the cosmological-constant problem. Recently, many
attempts have been undertaken to solve this paradox; 
see Refs.~\cite{w,wt}.  In this paper, we calculate a value 
of the cosmological constant which allows one to reconcile 
the above problem within a toy model being considered.

\bigskip
\bigskip
\noindent {\bf 2.\ Generalized Einstein equations}
\bigskip

We start by writing the generalized Einstein field equations for
$n$-dimensional spacetime,
\begin{equation}
R_{\mu\nu} - \frac{1}{2} g_{\mu\nu} R + \lambda g_{\mu\nu} =
\kappa T_{\mu\nu} \ ,
\label{1}
\end{equation}
where $R_{\mu\nu}$, $g_{\mu\nu}$ and $T_{\mu\nu}$, with $0\le
\mu,\nu\le (n - 1) \in {\bf N}$, are the components of the Ricci, 
metric and stress--energy tensors, respectively, and the quantity
$\lambda$ denotes the cosmological constant; one also defines
$R\equiv R_{\mu\nu} g^{\mu\nu}$ and $\kappa \equiv - 8 \pi G / c^4$
where the symbols $G$ and $c$ stand for the gravitational constant 
and for the speed of light in the vacuum, respectively. 
We assume as well that the relationship 
between the Ricci and the Riemann--Christoffel tensors is of the form
$R_{\mu\nu} \equiv {R^\alpha}_{\mu\nu\alpha} \equiv \partial_\nu
\Gamma_{\mu\alpha}^\alpha
- \partial_\alpha \Gamma^\alpha_{\mu\nu} +
\Gamma^\beta_{\mu\alpha} \Gamma^\alpha_{\beta\nu} - 
\Gamma^\beta_{\mu\nu} \Gamma^\alpha_{\beta\alpha}$ and the signature 
of the metric tensor $g_{\mu\nu}$ is equal to $(2 - n)$. 
We expect that the tensor $T_{\mu\nu}$ is given by the sum of 
two stress--energy tensors:\ that of the vacuum, 
$T_{\mu\nu}^{\it vac}$, and that of the ``ordinary'' matter and/or 
energy, $T_{\mu\nu}^{\it ord}$. The latter tensor should in principle 
be independent of the vacuum parameters.

\newpage
\noindent {\bf 3.\ Vacuum-energy density}
\bigskip

The stress--energy tensor $T_{\mu\nu}^{\it vac}$ for an empty space 
-- i.e., that of the vacuum -- should be of the form 
$T_{\mu\nu}^{\it vac} = \rho g_{\mu\nu}$ where $\rho$ denotes 
the (average) vacuum-energy density; see Ref.~\cite{w}.
On the basis of Eqs.~(\ref{1}) we then can 
assume that the observed, ``effective'' value $\lambda_{\it eff}$ 
of the cosmological constant reads
\begin{equation}
\lambda_{\it eff} = \lambda - \kappa \rho \ .
\label{2}
\end{equation}
In order to solve Eq.~(\ref{2}), we must first calculate the
quantity $\rho$ with the use of the ``standard'' quantum field
theory. For the vacuum, the expectation value of the Hamiltonian of 
a free scalar field with the mass $m$ in a flat (four-dimensional)
spacetime,
\begin{equation}
{\cal H} = \sum_{\bf k} \hbar \omega_{\bf k} \!\left( a_{\bf k}^+
a_{\bf k} + \frac{1}{2} \right) \ ,
\label{H}
\end{equation}
is given by the relationship
\begin{equation}
E = \langle 0 \vert {\cal H} \vert 0 \rangle = \frac{1}{2}
\sum_{\bf k}
\hbar \omega_{\bf k} \equiv \frac{1}{2} \sum_{\bf k} 
(-1)^\ell E_{\bf k}
\label{E}
\end{equation}
where $E_{\bf k} = (-1)^\ell (E_{\bf p}^2 + E_m^2)^{1/2}$ and 
$E_{\bf p} \equiv (-1)^\ell c |{\bf p}_{\bf k}| = (-1)^\ell
c \hbar |{\bf k}|$ as well as $E_m \equiv m c^2$; the parameter
$\ell$ is equal to $2$ or to $1$ for virtual particles or for 
virtual antiparticles, respectively, the quantity $\bf k$ 
signifies the wave vector of a homogeneous scalar plane 
wave, and the symbol $\hbar$ stands for the Planck constant.   
We assume that the sum over $\bf k$ in Eq.~(\ref{E})
should be performed up to a cutoff circular frequency
which we denote by $\omega_c$, so $0 \leq \omega_{\bf k}
\leq \omega_c$. The energy of the zero-point field with 
the (scalar) mass $m$ can be obtained as
a sum of the zero-point energies of all normal modes of this field
up to the cutoff energy equal to $\hbar \omega_c$.
Thus, in order to incorporate into
the sum in Eq.~(\ref{E}) all the possible zero-point fields, we
must perform the sum (or rather the integral) over all the possible
values -- of both the wave vector $\bf k$ and the 
mass $m$ -- for which the condition $\omega_{\bf k} \in 
[0, \omega_c]$ is fulfilled. To this end, we will {\it parametrize} 
the expressions for the energies $E_{\bf p}$ and $E_m$ with the use 
of two auxiliary variables which are $\phi \in [0, 2 \pi]$ and 
$K \equiv\pm\omega/c \in [- K_c, K_c]$ where $K_c = \omega_c/c$. 
The parametrization is accomplished by means of the substitution 
$E_{\bf p} = c\hbar K \cos \phi$ and $E_m = c\hbar 
K \sin \phi$, so the values of the quantities $E_{\bf p}$ and $E_m$ 
belong to the energy intervals $[- \hbar \omega_c, \hbar \omega_c]$ 
and $[0, \hbar \omega_c]$, respectively, which is attained
for the following set of the values of the auxiliary variables,
\begin{eqnarray}
{\cal I} (\phi, K) \!\!\! & \equiv & \!\!\! 
\left\{\mbox{$(\phi, K)$:} 
\right. \nonumber\\ \!\!\!  & \empty & \!\!\! \hspace{-0.619in}
\left. \left( \phi \in  \left[0, \frac{\pi}{2} \right] \; \& \;
K \in [0, K_c] \; \cup \; \phi \in \left[\pi, \frac{3\pi}{2} \right] 
\; \& \; K \in [- K_c, 0] \right) \; {\mbox{for $\ell = 2$}}
\right. \nonumber \\  \!\!\! & \hspace{-1.213in} \cup & \!\!\! 
\hspace{-0.665125in} \left.  \; \left(
\phi \in  \left[\frac{\pi}{2}, \pi \right] \; \& \; K \in [0, K_c] 
\; \cup \; \phi \in \left[\frac{3\pi}{2}, 2\pi \right] \; 
\& \; K \in [- K_c, 0] \right) \; {\mbox{for $\ell = 1$}}
\right\}. \nonumber
\end{eqnarray}

Assuming the four-dimensional ($1 + 3$) spacetime which is subject 
to our direct perception to be flat, one can now easily calculate 
the (average) vacuum-energy density $\rho$ for the case of
the vanishing effective Unruh--Davies temperature, i.e., 
in an inertial (Lorentz) reference frame,
\begin{equation}
\rho = \frac{1}{2} \int_{{\cal I} (\phi, K)} d\phi \, dK \, 
\frac{4\pi K^2}{(2\pi)^3} \sqrt{c^2 \hbar^2 K^2 \!\left(\cos^2 
\phi + \sin^2 \phi\right)} \, = \frac{c\hbar K_c^4}{8\pi} \ ;
\label{rho}
\end{equation}
cf.\ sections 7.2 and 7.8. To proceed futher, we have to estimate 
the value of the cutoff wave-number $K_c$ occurring in the formula 
(\ref{rho}). It should be stressed that here we do not assume
{\it a priori} any particular value for the quantity $K_c$; it can
fall within a very broad range of many orders of magnitude, which 
is not forbidden by experimental data. One can add as well that it 
seems to be impossible to extract any heat
energy from the average energy of the vacuum fluctuations occurring
at the absolute temperature equal to zero; see Ref.~\cite{Cole}
where also, among others, the correct form of the zero-point
energy spectrum of the electromagnetic thermal radiation (for which
one has $m = 0$) is deduced on the basis of the laws of the 
{\it classical} thermodynamics.

\bigskip
\bigskip
\noindent {\bf 4.\ A toy model of the spacetime}
\bigskip

In order to calculate the exact value of $K_c$, we should introduce
into our considerations a particular model concerning the topology
of a spacetime with the cosmological constant. In this paper we 
will use a toy model investigated in detail in Ref.~\cite{wt}. 
The toy model is based on a
five-dimensional spacetime which consists of the ordinary
four-dimensional spacetime and of an additional (macroscopically
unobservable) spatial dimension which acts as a carrier of the
vacuum pressure linked with the (negative) cosmological constant. 
The spacetime of the toy model investigated in Ref.~\cite{wt} is 
thus of the form ${\bf R}^1 (\mbox{\it time}) \times {\bf R}^1
(\mbox{\it extra spatial dimension}) \times {\bf R}^3 (\mbox{\it
three-dimensional space})$, and the vacuum stress--energy tensor is
defined to be $T_{\mu\nu}^{\it vac} = (\lambda_v/\kappa) 
g_{\mu\nu}$ for $\mu = 0, 1$ and $\nu = 0, 1$ or
$T_{\mu\nu}^{\it vac} = 0$ otherwise, 
with some appropriately adjusted value of $\lambda_v \in {\bf R}$. 
For the three-dimensional space ${\bf R}^3$ which is subject to our
direct perception, one then has $({\rm diag} \, T_{\mu\nu}^{\it vac}) 
= (\rho, p, 0, 0, 0)$ where the vacuum-energy density $\rho$ and 
the vacuum pressure $p$ obviously satisfy the vacuum equation of
state, $( \rho + p ) = 0$. It turns out that if we assume the 
diagonalized metric of the space ${\bf R}^3$ (i.e., the components
with $\mu = \nu = 2, 3, 4$ of the diagonalized metric tensor 
$g_{\mu\nu}$) to be independent of the coordinate $a$ of the 
additional spatial dimension, e.g., in order to avoid topological 
irregularities in the limits $a \to \pm \infty$ as well as any
instabilities, then we obtain that 
$\lambda_v = \lambda$ for any value of the quantity $\lambda$. 
Obviously, this result is equivalent to the requirement that 
$\lambda_{\it eff} = 0$ for $\mu, \nu = 0, 1$; note that such
a condition is formulated below in this paper, on a purely 
phenomenological basis; see section 5 and the end of section 6.

For instance, in the case of an empty
space, the solution of Eqs.~(\ref{1}) with the stress--energy tensor
$T_{\mu\nu}^{\it vac}$ defined above and for a negative value of 
the cosmological constant $\lambda$ reads
\begin{eqnarray}
ds^2 = \left(1 + |\lambda| a^2\right)\!  c^2 dt^2 - \left( 1 +
|\lambda| a^2\right)^{\! -1} da^2 - d x^2 - d y^2 - d z^2
\label{sol}
\end{eqnarray}
where $(ct, a, x, y, z) \equiv (x^0, x^1, x^2, x^3, x^4)$ and $a$
denotes the coordinate of the additional spatial dimension. 
It can easily be shown that no particle with a finite energy, 
which is moving in the spacetime described by Eq.~(\ref{sol}), 
could irrevocably leave the nearest neighbourhood $a \sim \pm \,
\pi |\lambda|^{-1/2}$ of the spacetime given by $a=0$; see 
Ref.~\cite{wt}. Note that assuming $(x, y, z) = {\it const}$, 
we obtain the metric which describes the covering surface 
${\bf R}^1 \times {\bf R}^1$ of the anti-de Sitter two-dimensional
spacetime ${\bf S}^1 \times {\bf R}^1$ with the (negative) 
cosmological constant $\lambda$ of a still unspecified value; such 
a spacetime contains ``global'' closed time-like curves with the 
retracing (coordinate) time equal 
to $T = 2\pi/(c |\lambda|^{1/2})$; see figure 1. 

In the presence of a spherically symmetric object with the rest mass 
$m$ in an empty (asymptotically flat) space, an exact solution of 
equations (\ref{1}) with the stress--energy tensor 
$T_{\mu\nu}^{\it vac}$ defined above takes the following form,
\begin{eqnarray}
ds^2 = \left(1 - \frac{2 M}{r} \right)\!  \left(1 + |\lambda|
a^2\right)\!  c^2 dt^2 - \left( 1 + |\lambda| a^2\right)^{\! -1} da^2
\nonumber\\
- \left( 1 - \frac{2 M}{r} \right)^{\! -1} d r^2 - r^2 d \theta^2 -
r^2 \sin^2 \theta \, d \varphi^2 \ ,
\label{schw}
\end{eqnarray}  
where $r\in (0, \infty)$, $\theta\in [0, \pi]$ as well as 
$\varphi\in [0, 2\pi)$ are the polar coordinates of the 
three-dimensional space $(x, y, z)$ around the mass $m$, and the 
quantity $M$ is given by the expression $M \equiv G m/c^2$.
Note that in the case of the metric (\ref{schw})
the energy $E$ of a free test (point-like) particle with a finite 
rest mass $m_0$ reads
\begin{eqnarray}
E^2 = c^2 \!\left( 1 - \frac{2 M}{r} \right) 
\!\left[ m_0^2 c^2 \right. \nonumber \\ \left.
+ \left( 1 - \frac{2 M}{r} \right)^{\! -1}
\!\left(p^r\right)^2
+ r^2 \!\left( p^\theta \right)^2 + r^2 \!\left( \sin^2 \theta 
\right) \!\left( p^\varphi \right)^2\right]
\label{schw_enr}
\end{eqnarray}
where we define $p^r \equiv m_0 dr/d\tau$, $p^\theta 
\equiv m_0 d\theta/d\tau$ as well as $p^\varphi \equiv 
m_0 d\varphi/d\tau$, and the symbol $\tau$ denotes
the proper time of the particle being considered; the quantity $E$ 
is defined here in the same form as the particle's energy in the 
four-dimensional spacetime ${\bf R}^1 \times {\bf R}^3$, i.e., 
in the absence of the additional spatial dimension $a$.

In turn, the cosmological solution to Eqs.~(\ref{1}) -- for the 
Universe being spatially homogeneous and isotropic with respect 
to the three dimensions $(x, y, z)$ -- is given by the metric
\begin{eqnarray}
ds^2 = \left(1 + |\lambda| a^2\right)\!  c^2 dt^2 - \left( 1 +
|\lambda| a^2\right)^{\! -1} da^2 
\nonumber \\ - [R(t)]^2 \!\left[
\!\left( 1 - k r^2 \right)^{\! -1} d r^2 + r^2 d \theta^2 +
r^2 \sin^2 \theta \, d \varphi^2 \right]
\label{sol_cos}
\end{eqnarray}
where $r\in [0, \infty)$, $\theta\in [0, \pi]$ as well as
$\varphi \in [0, 2\pi)$ are the dimensionless polar ``comoving'' 
coordinates and $R(t)$ denotes the expansion parameter 
(or the cosmic scale factor) which satisfies the field
equations resulting from Eqs.~(\ref{1}) and from the specific 
form of the stress--energy tensor $T_{\mu\nu}^{\it ord}$. Note
that the self-consistency of the assumed model requires
the constant curvature parameter $k$ to be equal to zero in 
the metric (\ref{sol_cos}); see Ref.~\cite{wt} for details.

The geometric arguments resulting from the toy model 
(see figure 1) lead one to the conclusion that the cutoff time 
$T_c$ is given by the relationship $T_c = 2\pi/(c |\lambda|^{1/2})$,
so the cutoff length is equal to $L_c = c \, T_c =
2\pi/|\lambda|^{1/2}$ and the cutoff energy -- to $E_c =
h/T_c = c\hbar |\lambda|^{1/2}$ where 
$h \equiv 2 \pi \hbar$.  We thus have $K_c = |\lambda|^{1/2}$,
since $K_c \equiv \omega_c/c = E_c/(c \hbar)$.
It is interesting to note that the above formula can also be 
obtained with the use of the Bohr--Sommerfeld quantization condition 
for a particle with the cutoff energy, which oscillates with the 
velocity $c$ in the $(ct)$-direction of the anti-de Sitter spacetime 
of the toy model (see figure 1),
\begin{equation}
\oint \hbar k_c \, d(ct) = h \qquad \Longrightarrow \qquad 
c k_c T_c = 2 \pi \ ,
\label{BS}
\end{equation}
where $k_c$ denotes the cutoff length of the wave vector 
oriented along the positive axis in the above-mentioned direction;
we see that the dispersion relation $\omega_c = c k_c$ holds obviously
for the oscillations considered here, and the relationship between 
the cutoff wave-number $k_c$ and the cutoff energy $E_c$ is given by 
the formula $k_c = E_c/(c \hbar)$, since one has $E_c = \hbar \omega_c$. 
The geometric properties of the toy-model spacetime imply that 
$T_c = 2 \pi/(c |\lambda|^{1/2})$, so indeed we have 
$K_c = k_c = |\lambda|^{1/2}$; additionally, one 
obtains the equality $E_c = c \hbar |\lambda|^{1/2}$.

\bigskip
\bigskip
\noindent {\bf 5.\ Astronomical observations}
\bigskip

Thus, the vacuum-energy density calculated in the formula 
(\ref{rho}) reads
\begin{equation}
\rho = \frac{c \hbar \lambda^2}{8 \pi} \ .
\label{rho2}
\end{equation}
In turn, taking into account various astronomical observations
and their interpretation within the toy model, 
we can put $\lambda_{\it eff} = 0$ in Eq.~(\ref{2}); for instance, 
the data coming from the observations
of distant supernovae and suggesting a small, but non-zero value
of the observed cosmological constant $\lambda_{\it eff}$ can be
interpreted and explained by maintaining the condition 
$\lambda_{\it eff} = 0$ unchanged, simultaneously with the 
introduction of some local inhomogeneities into the Hubble-constant 
field; it is clear that such inhomogeneities may occur, e.g., due to 
peculiar streaming motions towards the regions of space where 
especially large amounts of mass are concentrated -- just as it happens
in the case of the Great Attractor that is situated at the redshift 
$z \sim 0.02$; see Ref.~\cite{wt} and references therein. 
An interesting experimental 
test concerning the possible vanishing of the value of the observed 
cosmological constant $\lambda_{\it eff}$ could come from an 
E\"otv\"os-type experiment performed for masses made from various 
materials (i.e., from aluminum and from a monel metal like copper 
or silver); see Refs.~\cite{Ross,Gillies}.

Here it should be remarked that for the vacuum stress--energy tensor 
$T_{\mu\nu}^{\it vac}$ defined in section 4 (which is non-vanishing 
only when $\mu, \nu = 0, 1$), a formula for $\lambda_{\it eff}$ 
takes the form given by Eq.~(\ref{2}) only for the components with 
$\mu = 0, 1$ and $\nu = 0, 1$ of the Einstein equations (\ref{1}); 
otherwise, one obtains that $\lambda_{\it eff} = \lambda$, 
since for $\mu \neq 0, 1$ or $\nu \neq 0, 1$ 
we have by definition $T_{\mu\nu}^{\it vac} = 0$. 
In the latter case, however, if we assume that 
$\lambda_{\it eff} = 0$ (so also $\lambda = 0$), then we arrive at the 
conclusions which are not consistent with observational data:\ for 
instance, the vacuum-energy density given by expression (\ref{rho2}) 
would be equal to zero, and the metric (\ref{sol}) with $\lambda = 0$ 
would not prefer in any way the three-dimensional space $a = 0$ within
which and around which -- according to the toy model -- the whole 
matter and radiation present in the Universe should be concentrated.
Thus, let us demand the condition $\lambda_{\it eff} = 0$ to be 
fulfilled for the $(t,t)$-component of the Einstein equations 
(\ref{1}); as a confirmation of the correctness of the above
requirement we can note the fact that in the case of an empty space and
sufficiently short distances the component with $\mu = 0$ and 
$\nu = 0$ of the Einstein equations should approach 
the generalized Poisson equation,
\begin{equation} 
\nabla^2_{\bf r} V ({\bf r}) = \frac{4 \pi G}{c^2} 
\, {\rm Tr} ({\rm diag} \, T_{\mu\nu}) = 0
\label{Poisson}
\end{equation}
for $a = 0$ and with the symbol $\bf r$ denoting the position vector
${\bf r} \equiv [ x, y, z ]$ as well as with the function $V = V 
({\bf r})$ being the gravitational potential, and equation 
(\ref{Poisson}) is actually consistent with the condition that in an 
empty three-dimensional space being subject to our direct perception
the observed (effective) cosmological constant $\lambda_{\it eff}$ 
is equal to zero. 

\bigskip
\bigskip
\noindent {\bf 6.\ A value for the cosmological constant}
\bigskip

Substituting expression (\ref{rho2}) into 
Eq.~(\ref{2}) and taking $\lambda_{\it eff} = 0$, we then come to 
the conclusion that the value of the cosmological
constant is given by the combination of the fundamental constants
of nature,
\begin{equation}
\lambda = - \frac{c^3}{G \hbar} \cong - 3.829 \times 10^{69}
\;\, {\rm m^{-2}} \ .
\label{cc}
\end{equation}
For such a value of the cosmological constant, 
the cutoff time $T_c$ is of the order of the Planck time 
$T_{\it Pl} \sim 10^{-43}$ s, so the cutoff length $L_c$ is of the 
order of the Planck length $L_{\it Pl} \sim 10^{-34}$ m and the 
cutoff energy $E_c$ is of the order of the Planck energy 
$E_{\it Pl} \sim 10^{28}$ eV $\sim 10^9$ J, as one might have 
expected. Namely, one 
intuitively assumes that it is not possible for any field to propagate 
at energies for which (generalized) Compton wavelengths 
defined as $h/(m_{\it eq} c) = 
2 \pi c/\omega$ are less than the Schwarzschild diameter 
$4 G m_{\it eq}/c^2$ \cite{Strom}; 
such a situation would happen, for instance, in the case of 
both massive and massless particles 
with ``equivalent masses'' $m_{\it eq} \equiv \hbar \omega/c^2$ above
the Planck mass (multiplied by a factor of $\pi^{1/2}/2^{1/2}$), 
or with generalized Compton wavelengths below the Planck length times
the constant $(2/\pi)^{1/2}$. Such particles would simply possess an
event horizon formed around them, so they would disconnect from the
surrounding spacetime; see also section 7.6 of this paper. Thus, 
the Planck energy $E_{\it Pl}$ should
provide a natural cutoff for energies of normal modes of any field,
and this is actually confirmed by the above calculations. In fact, 
an important conclusion of this paper can be formulated in such a way
that the vanishing of the observed value $\lambda_{\it eff}$ of the 
cosmological constant is in the toy model a consequence of the existence 
of an energy cutoff at the level of the Planck scale $E_{\it Pl}$. 
It is easy to show that if we require the condition $|\lambda_{\it eff}| 
\le 10^{-50}$ m$^{-2}$ to be fulfilled, then the relative departure 
of the cutoff energy $E_c = c \hbar |\lambda|^{1/2}$ from the Planck 
energy $E_{\it Pl} \equiv (c^5 \hbar/G)^{1/2}$ should not be greater 
than a factor of the order of $10^{-120}$, i.e., one obtains that
$|E_c - E_{\it Pl}|/E_{\it Pl} \le 10^{-120}$.

Note that exactly the same value of the cosmological constant as
calculated above was {\it postulated} in Ref.~\cite{wt} to give 
results concerning the motion
of a free test particle, which are in agreement with experimental
data. In particular, with the use of Eq.~(\ref{cc}) one is able to
recover the appropriate proportionality constant in the equation of
the motion of a free test particle which is travelling in the spacetime
of the toy model described -- in the case of an empty space -- by the 
metric given by expression (\ref{sol}); as a by-product, one obtains 
the Planck--Einstein formula $E = \hbar \omega$ which, however, assumes
therein a completely new -- geometric and entirely classical -- 
meaning; see also, for instance, sections 7.1 and 7.2 of this paper. 
Similarly, in the case of the motion of a free test particle in the 
spacetime with the metric (\ref{schw}), one obtains that $E = \hbar 
\omega (1 - 2 M/r)^{1/2}$ where the energy $E$ is now given by 
expression (\ref{schw_enr}). It should be stressed here that the 
relationship between the particle's energy $E$ and the circular
frequency $\omega$ of the particle's oscillations in the additional 
spatial dimension of the spacetime described by the metric (\ref{sol}) 
is given by the formula $E = \hbar \omega$ only if we demand the 
condition $\lambda_v = \lambda$ to be fulfilled for the vacuum 
stress--energy tensor $T_{\mu\nu}^{\it vac}$ which is defined in 
section 4. The assumption that the difference $(\lambda_v - \lambda)$ 
remains a non-vanishing quantity would imply that the equality $E = 
\hbar \omega$ is not an exact formula in the case of the motion of a 
free test particle travelling in the spacetime with the metric 
(\ref{sol}); note that a possible deviation from the equality 
$E = \hbar \omega$ could in principle be detected by some 
sufficiently sensitive experiments and/or by appropriate 
astronomical observations. Obviously, analogical conclusions hold
in the case of a test particle moving in the spacetime with the metric
(\ref{schw}) and for the detection of a possible deviation from the
formula $E = \hbar \omega (1 - 2 M/r)^{1/2}$ which expresses there 
the particle's energy.

\bigskip
\bigskip
\noindent {\bf 7.\ A model of a particle--antiparticle pair and
its consequences}

\bigskip
\bigskip
\noindent {\it 7.1.\ Introduction}
\bigskip

The integrated equation of motion for a test particle with the 
rest mass $m$, which is moving in the spacetime described by the 
metric (\ref{sol}) reads
\begin{eqnarray}
U^2 = \left( 1 + |\lambda| a^2 \right) \!\left( m^2 c^4 +
c^2 {\bf p}^2 \right) + c^2 \!\left( p^a \right)^2 \nonumber \\
\equiv \left( 1 + |\lambda| a^2 \right) E^2 
+ c^2 \!\left( p^a \right)^2
\label{enr}
\end{eqnarray}
where the quantities $U$, $m$ and 
${\bf p}$ are the particle's total energy, rest mass
and three-dimensional momentum vector ${\bf p} \equiv
[p^x, p^y, p^z] \equiv [p^2, p^3, p^4]$ in the
spacetime $a = 0$, respectively, and $c$ denotes 
the velocity of a massless particle with regard to the 
spacetime $a = 0$. In turn, one has $U \equiv c p_t \equiv c p_0$, 
${\bf p}^2 \equiv - \sum_{i=2}^4 p_i p^i = \sum_{i=2}^4 (p^i)^2$, 
and $p^a \equiv p^1 = da/d(\tau/m)$ with $\tau$ denoting the proper 
time of the particle; note that in the limit of a massless 
particle, the quantity $\tau/m$ remains finite and is still an 
affine parameter. The total energy $U$ of the particle, which
is a {\it hidden} parameter, remains constant for objects 
travelling along the geodesic lines, since the metric 
(\ref{sol}) is independent of the (coordinate) time $t$.

It is easy to see that a particle moving in the spacetime 
$a \cong 0$ is confined inside a potential well, proportional 
to the factor $(1 + |\lambda| a^2)^{1/2}$, which is enormously 
narrow along the additional spatial dimension $a$, but flat 
along the three ``physical'' ones $(x, y, z)$. Thus, the 
energy $E = c p_t (a = 0, p^a = 0)$ equal to the particle's 
total energy in the ``ordinary'' four-dimensional 
spacetime ${\bf R}^1 \times {\bf R}^3$ (i.e., in the absence of 
the additional spatial dimension $a$) can fluctuate to a very small 
extent only, and in the non-relativistic approach $p \equiv |{\bf p}| 
\ll m c$ assumed for the case of a massive particle one obtains easily 
the uncertainty relationship of purely classical origin, 
$\Delta t \, \Delta E \ge \hbar/2$; see Ref.~\cite{wt}.
We can note that if the spacetime of the toy model were 
four-dimensional, or special-relativistic (i.e., without an 
additional spatial dimension), then the uncertainty relationship 
for a toy-model massive particle would have the form 
$\Delta t \, \Delta E \ge 0$, which is manifestly wrong.

\bigskip
\bigskip
\noindent {\it 7.2.\ The model}
\bigskip

It is now worth realizing a very simple picture of the vacuum
fluctuations, which is possible within the toy model; namely,
each point of the spacetime $a = 0$ that is embedded in the 
five-dimensional manifold with the metric (\ref{sol})
can be represented as a superposition of two vibrations, or
excitations, each of which satisfies the integrated equation of 
motion (\ref{enr}). One can prove easily that the coordinates of 
such two vibrations occurring in the additional spatial dimension 
are given by
\begin{equation}
a_1 (\tau) = \sqrt{\frac{U^2 - E^2}{|\lambda| E^2}} \, 
\sin (\omega \tau)
\label{mot1}
\end{equation}
and
\begin{equation}
a_2 (\tau) = \sqrt{\frac{U^2 - E^2}{|\lambda| E^2}} \, 
\sin (- \omega \tau)
\label{mot2}
\end{equation}
where one defines
$\omega \equiv \pm E/\hbar = (m^2 c^4 + c^2 {\bf p}^2)^{1/2}/\hbar$,
with the signs ``$+$'' or ``$-$'' referring to the formulae 
(\ref{mot1}) or (\ref{mot2}), respectively, and the quantity $\tau$ 
denotes here an affine parameter in general, e.g., the proper time 
in the case of massive excitations; note that, strictly speaking, 
the argument of the sine function appearing in expressions
(\ref{mot1}) or (\ref{mot2}) is respectively equal
to $\pm \omega (m_c \tau/m)$ rather than to $\pm \omega \tau$, 
where $m_c \equiv E_c/c^2 = \hbar |\lambda|^{1/2}/c$.
One then can identify those two vibrations with a probe virtual
particle and its antiparticle remaining at rest or moving together 
-- connected, but non-interacting -- with a uniform rectilinear motion 
in regard to an inertial reference frame of the spacetime $a = 0$ 
which is embedded in the five-dimensional manifold with the metric 
(\ref{sol}); see also Ref.~\cite{wt}. We suppose that the two
excitations being considered do not interact with each other through 
any forces, since the non-gravitational interactions are expected 
to act and propagate in the spacetime $a \cong 0$ only; see, however,
section 7.8. It is worth adding that in the case of the metric 
(\ref{schw}) the relationships (\ref{mot1}) and (\ref{mot2}) hold
as well, with the circular frequency $\omega$ defined now as 
$\omega \equiv \pm (E/\hbar) (1 - 2 M/r)^{-1/2}$, the energy $E$ given 
by expression (\ref{schw_enr}) and the quantity $M$ denoting the mass 
of a (real) spherically symmetric object. Note that in
the formula (\ref{mot1}) and in the rest of this section we designate 
various quantities concerning the particle -- both a virtual and
a real one -- by adding the suffix ``$1$'', and we label those 
same quantities characterizing the corresponding antiparticle -- 
by inserting the suffix ``$2$'', just as in expression (\ref{mot2}). 

Obviously, the considerations contained in the above paragraph
are easy to be generalized; namely, each point of the spacetime $a = 0$
can actually be represented as a superposition of an arbitrary (e.g.,
infinite) number/amount of the pairs of excitations described by 
the formulae (\ref{mot1}) and (\ref{mot2}), with taking into account
the allowed values of the mass $m$ as well as lengths and directions 
of the momentum vector $\bf p$; see section 3 where, according to 
the model here-presented, the vacuum-energy density $\rho$ should
result from the superposition of {\it all} the possible 
(allowed) pairs of vibrations. Clearly, all the above-mentioned pairs 
of excitations can additionally differ among themselves in values of 
the hidden parameters $U$ (which are obviously the same for both 
the constituent vibrations of each pair); see also Ref.~\cite{wt}.
Bearing in mind the form of expressions (\ref{mot1}) and (\ref{mot2}),
it is then easy to conclude that the values of both the vacuum-energy 
density $\rho$ calculated in the formula (\ref{rho}) and the cutoff 
circular frequency $\omega_c$, or wave-number $K_c$ determined in 
section 4 of this paper do not depend on the particular choice of an 
(inertial) reference frame of the spacetime $a = 0$, which means that 
the above quantities are Lorentz-invariant.

Since for a virtual particle and for its antiparticle there hold
the relationships $a_1 (\tau) \propto \sin (\omega \tau)$ and
$a_2 (\tau) \propto \sin(- \omega \tau)$, respectively, so the 
energies $E$ of the virtual particle and of its antiparticle 
should satisfy the equalities $E_1 = \hbar \omega$ and 
$E_2 = - \hbar \omega$, respectively; thus,
the total energy $E_{1 2} \equiv E_1 + E_2$ of the virtual
particle--antiparticle pair is equal to zero. This corresponds
to the fact that the average vibration representing the virtual
particle--antiparticle pair vanishes, as one has $a_{1 2} (\tau)
\equiv a_1 (\tau) + a_2 (\tau) = 0$ for any value of the affine
parameter $\tau$. Simultaneously, each virtual
particle--antiparticle excitation gives a non-zero contribution
to the vacuum energy defined as in Eq.~(\ref{E}), since
the average absolute value of excitation, which is given 
by the quantity ${\bar a}_{1 2} (\tau) \equiv 
(| a_1 (\tau)| + | a_2 (\tau)|)/2$, obviously remains (almost) 
always positive for any virtual particle--antiparticle vibration.
It is then worth adding that a factor of $1/2$, appearing
in a purely classical context in the above definition of the 
function ${\bar a}_{1 2}$, corresponds clearly to exactly the same
factor occurring in Eq.~(\ref{E}), which is therein, however, entirely
of formal-quantum origin. In other words, one has ${\bar a}_{1 2} 
(\tau) \propto \sin[(|E_1| + |E_2|) \tau/(2 \hbar)]$, which means 
that the contribution to the vacuum energy coming from 
a virtual particle or its antiparticle separately
is equal to $|E_1|/2$ or to $|E_2|/2$, respectively, so it is 
exactly the same as in the case of the quantum-mechanical harmonic 
oscillator (with the frequencies $|E_1|/\hbar$ or $|E_2|/\hbar$,
respectively) remaining in its ground energy-state; note that
the above conclusion holds also for the root-mean-square
value of virtual particle--antiparticle excitation, which is
defined as ${\widehat{a}}_{1 2} (\tau) \equiv ([a_1^2 (\tau) + 
a_2^2 (\tau)]/2)^{1/2}$.

In turn, one can easily note that because of the existence of the 
energy cutoff in the toy model, the (possible) values of 
the circular frequency $\omega$ of the particle's vibration should 
for any particle (or antiparticle) belong to the interval $\omega 
\in [0, \omega_c]$, which implies the limits on the value of the 
particle's rest mass $m \in [0, \hbar\omega_c/c^2]$ as well as on 
the values of the particle's momentum $p \in [0, \hbar \omega_c/c]$.

\bigskip
\bigskip
\noindent {\it 7.3.\ A virtual particle--antiparticle pair}
\bigskip

Let us now consider a pair of a virtual particle and its
antiparticle. According to the energy definitions following 
Eq.~(\ref{E}), for a virtual particle and for its antiparticle 
separately we have
\begin{eqnarray}
E_{{\bf k} 1} = \hbar \omega \ , \qquad E_{{\bf p} 1} = c p \qquad
{\mbox {\rm and}} \qquad E_{m 1} = m c^2 
\label{vir1}
\end{eqnarray} 
as well as
\begin{eqnarray}
E_{{\bf k} 2} = - \hbar \omega \ , \qquad E_{{\bf p} 2} = - c p 
\qquad {\mbox {\rm and}} \qquad E_{m 2} = m c^2 \ ,
\label{vir2}
\end{eqnarray}
respectively. Thus, for a {\it virtual} particle--antiparticle 
pair, taken as a whole, one obtains that
\begin{eqnarray}
E_{{\bf k} 1 2} = 0 \ , \qquad E_{{\bf p} 1 2} = 0 \qquad
{\mbox {\rm and}} \qquad E_{m 1 2} = 2 m c^2 \ . 
\label{vir3}
\end{eqnarray} 
In turn, from the integrated equation of motion (\ref{enr}) 
for the virtual particle--antiparticle pair it results that
\begin{equation}
U_{1 2} = 2 m c^2 \ ,
\label{tog}
\end{equation}
since $p^a_{1 2} \equiv p^1_{1 2} = 0$, as we have 
$a_{1 2} (\tau) = 0$ for any value of the affine parameter $\tau$.

It then seems that the above picture of the pairs of virtual 
particles and their antiparticles can justify the summation
(or rather the integral) over the continuous mass $m$ spectrum, 
which is performed while calculating the
vacuum-energy density $\rho$ in Eq.~(\ref{rho}), since the mass $m$
appearing in expressions (\ref{vir1})--(\ref{tog}) can assume any 
value from the interval $[0, \hbar \omega_c/c^2]$. In turn, according 
to the equality (\ref{tog}), one could speculate as well that the 
fluctuations of the vacuum might be at the origin of the physical
concept of {\it mass}; see also, e.g., sections 7.6 and 7.8 of this
paper as well as Ref.~\cite{Alicki} where one considers the possible 
electromagnetic nature of particle's mass in the context of the 
generalized Maxwell equations. For instance, on the basis of 
the formula (\ref{tog}) we can say that the 
spacetime $a = 0$ is filled with the enormously dense field of 
virtual mass. On the other hand, the relationship (\ref{tog}) 
suggests as well that the spacetime $a = 0$ is filled with the huge 
amount of the hidden energy $U$ whose source would most probably be 
situated beyond the observable Universe given by $a \cong 0$.

It is also worth adding that the virtual particle--antiparticle 
excitation can last for an arbitrarily long time (i.e., one has
$\Delta t \to \infty$), since the total energy $E_{{\bf k} 1 2}$ 
of the virtual particle--antiparticle pair is equal to zero and 
the uncertainty relation $\Delta t \, \Delta E \ge \hbar/2$
holds. Taking into account the three uncertainty relationships, 
$\Delta x^i \Delta p^i \ge \hbar/2$ for $i = 2, 3, 4$, which are
satisfied in the toy model as well \cite{wt}, we come to the
conclusion that the vibrations being considered fill the whole 
spacetime given by $a = 0$, as one has $\Delta p^i_{1 2} = 0$ 
for $i = 2, 3, 4$ and $\Delta E_{{\bf k} 1 2} = 0$. Similarly, it 
should be remarked here that in the spacetime described 
by the metric (\ref{schw}) the uncertainty relation can be written 
as $\Delta \tau \, \Delta E \ge (\hbar/2) (1 - 2 M/r)^{1/2}$, since
for such a case we obtain that $\Delta \tau \, \Delta \omega \ge
1/2$ and $\Delta E = \hbar (1 - 2 M/r)^{1/2} \, \Delta \omega$
where the energy $E$ is given by expression (\ref{schw_enr}); see
Ref.~\cite{wt}. Thus, the relationship $\Delta \tau \, \Delta E
\ge 0$ is satisfied in the limit $r \to (2 M)^+$, i.e., on the 
event horizon of a Schwarzschild black hole, which is evidently
consistent with the condition $\tau = {\it const}$ that holds
for $r = 2 M = {\it const}$ and $d \theta = 0 = d \varphi$ assumed
in the metric (\ref{schw}) with $a = 0$. Obviously, in such a case 
one still has the relation $\Delta t \, \Delta E \ge \hbar/2$ 
fulfilled for the coordinate time $t$.

\newpage
\noindent {\it 7.4.\ Creation of a real photon}
\bigskip

Let us now imagine that the portion of energy equal to 
$E = 2 \hbar \omega$, with the momentum vector assumed
to be equal to $2 {\widetilde{\bf p}}$,
is supplied to a virtual particle--antiparticle pair. As an example
of such a process we can consider the spontaneous emission, where 
an atom undergoes the transition from an excited initial state to a 
lower-energy, final state, simultaneously emitting a photon with the
energy equal to the difference between the energies of the excited 
and of the final state of the atom. It is commonly known that the
spontaneous-emission process can actually be regarded as an emission
process induced by the fluctuations of the vacuum. Let us assume that
the energy $E = 2 \hbar \omega$ of the transition between the excited
and the lower-energy state of the atom is absorbed by one of the 
vacuum (virtual particle--antiparticle) excitations described above 
in sections 7.2 and 7.3. We suppose that after such an absorption 
one has for the vibrations being considered,
\begin{eqnarray}
E_{{\bf k} 1} = \hbar \omega \ , \qquad E_{{\bf p} 1} = c p \qquad
{\mbox {\rm and}} \qquad E_{m 1} = m c^2 
\label{phot1}
\end{eqnarray}
as well as 
\begin{eqnarray}
E_{{\bf k} 2} = \hbar \omega \ , \qquad E_{{\bf p} 2} = c p \qquad
{\mbox {\rm and}} \qquad E_{m 2} = m c^2 \ , 
\label{phot2}
\end{eqnarray}
and both the excitations now form one {\it real} particle with
\begin{eqnarray}
E_{{\bf k} 1 2} = 2 \hbar \omega \ , \qquad E_{{\bf p} 1 2} = 
2 c {\widetilde{p}} \qquad {\mbox {\rm and}} \qquad E_{m 1 2} = 0 \ , 
\label{phot3}
\end{eqnarray}
which can be identified with a photon with the momentum vector
$2 {\widetilde{\bf p}}$ 
and with the energy $E = 2 \hbar \omega = 2 c {\widetilde{p}}$ 
equal to the energy of the atom's transition from the initial 
to the final energy state; obviously, the condition
$({\widetilde{p}})^2 = p^2 + (m c)^2$ should be fulfilled -- also in 
the case of the spacetime described by the metric (\ref{schw}), where 
one defines the momentum $p$ (or, analogically, ${\widetilde{p}}$) 
and the above-given energy $E$ as $p \equiv (- \sum_i g_{ii} p^i 
p^i)^{1/2}$ for $i = r, \theta, \varphi$ and $E^{(*)} \equiv c p^t 
(a = 0, p^a = 0) = (1 - 2 M/r)^{- 1/2} c p_t (a = 0, p^a = 0)$, 
respectively. We assume that the hidden energy $U$ of the photon 
contains (includes) the quantity $U_{1 2} = 2 m c^2$, as in the case 
of the virtual particle--antiparticle pair described in section 7.3; 
note that the energy $U_{1 2}$ gives no contribution to the (vanishing) 
rest energy $E_{m 1 2}$ of the photon. 

We should stress here that one can also consider real particles 
consisting of more than two vibrations; obviously, a real photon 
might be formed from two or more virtual particle--antiparticle 
pairs, each of which is excited -- in the way described above --
due to the absorption of one or
more energy quanta with appropriate values of energy $E$ and 
with relevant lengths and directions of momentum vector $\bf p$.
It is worth adding as well that the created photon satisfies
the wave equation which seems to be of a phenomenological rather 
than of a fundamental nature in view of the toy model being considered; 
see Ref.~\cite{wt}. We also would like to mention here that an
interesting model of a photon and of a neutrino formed as
extended compact particles has been proposed
in Ref.~\cite{Hunter}. It might be instructive to generalize the
considerations contained in that reference to the case of the
five-dimensional spacetime of the toy model.

From the above picture, it is clear that each photon consists of
the two components, or of the two vibration modes; presumably, 
an electromagnetic wave of the form
$({\bf E} + i c {\bf B})/2$ can be assigned to each of those two 
modes; here the quantities $\bf E$ and $\bf B$ are the strength 
vectors, respectively, of an electric and a magnetic field 
associated with a photon, and $i$ denotes the imaginary unit; 
see also section 7.8.

\bigskip
\bigskip
\noindent {\it 7.5.\ Creation of a real particle--antiparticle pair}
\bigskip

Now let us consider another experiment:\ One or more energy quanta
(e.g., photons) with a sufficiently large value of the total energy 
$E = 2 \hbar \omega$ and with the momentum vector equal to 
$2 {\widetilde{\bf p}}$ are absorbed by the vacuum excitation 
representing a virtual particle--antiparticle pair. As a consequence, 
a real particle--antiparticle pair is coming into existence. 
The {\it real} particle and its antiparticle possess the energies
\begin{eqnarray}
E_{{\bf k} 1} = \hbar \omega \ , \qquad E_{{\bf p} 1} = c p \qquad
{\mbox {\rm and}} \qquad E_{m 1} = m_0 c^2  
\label{par1}
\end{eqnarray}
as well as
\begin{eqnarray}
E_{{\bf k} 2} = \hbar \omega \ , \qquad E_{{\bf p} 2} = c p \qquad
{\mbox {\rm and}} \qquad E_{m 2} = m_0 c^2 \ ,
\label{par2}
\end{eqnarray}
respectively. Thus, we assume here that the energy $U_{1 2} =
2 m_0 c^2$, which is discussed also in the previous two subsections, 
splits into two components $U_1$ and $U_2$, each 
of which takes the value equal to $m_0 c^2$. Subsequently, in 
the process of a particle--antiparticle creation, the energies $U_1$
and $U_2$ undergo the transition to give non-vanishing contributions
into both the rest energies $E_{m 1}$ and $E_{m 2}$ of the created 
particle and of its antiparticle, respectively, so the virtual masses 
become the real ones. 
Obviously, the values of the energies $E_{\bf k}$, $E_{\bf p}$ 
and $E_m$ of both the created (real) particle and its antiparticle 
remain positive; see Ref.~\cite{Cost}. Additionally, both the created
particles may possess electric charge and spin, each of which is 
expected to have the same absolute value and opposite signs
for the particle and its antiparticle; see section 7.6. Note that each 
of the created particles satisfies the (phenomenological) Klein--Gordon 
equation; see Ref.~\cite{wt}.

Of course, the energies $E_{{\bf k} 1}$ and $E_{{\bf p} 1}$ 
in expression (\ref{par1}) could differ numerically from the energies 
$E_{{\bf k} 2}$ and $E_{{\bf p} 2}$ in expression (\ref{par2}), 
according to the laws of conservation of energy $E$ as well as
of momentum vector $\bf p$. In turn, 
the symbol $m$ with the suffix ``$0$'' denotes now the rest 
mass of an actually existing real particle, such as an electron, or 
a proton; obviously, there must exist some -- so far unknown -- 
``weak'' and/or ``strong'' stability condition(s) which
determine the values of the masses of the created 
particles:\ electron/positron, meson (its quarks)/antimeson (its 
antiquarks), proton (its quarks)/antiproton (its antiquarks), 
{\it etc.} Those particles, whose masses do not fulfil even the 
``weak'' stability condition, can also be created, but they are
expected to live for a {\it very} short time only; some of such
particles are known as resonances.

The question arises here as to whether the particle's oscillatory 
motion with respect to the additional spatial dimension $a$ can 
correspond to an effect which is known as the {\it Zitterbewegung} 
of a massive particle. The velocity of the {\it Zitterbewegung} 
should be equal to the speed of light $\pm c$, so we can demand the
condition
\begin{equation}
{\dot{a}}_{\it extr} (\tau) = \pm \lim_{v \to c^-} v = \pm c
\label{zitter}
\end{equation}
to be fulfilled for Eq.~(\ref{mot1}), or for Eq.~(\ref{mot2}) in the 
case of a massive antiparticle. Obviously, the proper velocity 
${\dot{a}} (\tau) \equiv d a(\tau)/d\tau$ assumes its extreme values 
for the moments of the proper time $\tau$ such that $a (\tau) = 0$, and 
the signs ``$\pm$'' of the proper velocity $\dot{a}$
in Eq.~(\ref{zitter}) occur alternately after each other with the 
lapse of the proper time $\tau$, because of the presence of the 
sine function in the formula (\ref{mot1}). In turn, the condition
$a = 0$ determines the spacetime where the Dirac equation holds, 
whose velocity eigenvalues are always equal exactly to $\pm c$. Thus,
the requirement (\ref{zitter}) is consistent with both the toy model
and the quantum theory.  Solving Eq.~(\ref{zitter}), we obtain 
that $(U^2 - E^2 ) = E^2_c = E^2_{\it Pl}$, so one has $U_{\it min} = 
E_{\it Pl}$, which seems to be a reasonable result stating that
the hidden energy $U$ of a free and single massive particle can 
never be less than the maximum energy $E$ which is attainable for 
a massive as well as for a massless particle in the toy model.
It is important to add that according to Eq.~(\ref{enr}) the 
above-described oscillatory motion of a massive particle with respect 
to the additional spatial dimension, which occurs in the toy model,
does not violate the principles of relativity, although the maximum 
velocity of such a motion can be equal to the speed of light $\pm c$ at
the moments of (the proper) time when the particle crosses the spacetime 
$a = 0$; see also Refs.~\cite{Dirac,Elbaz}. One should note as well 
that the rest energy $E_0 = m_0 c^2$  of a particle with the rest mass 
$m_0$ can be regarded as the kinetic energy of the particle's 
{\it Zitterbewegung}, since the condition $E_0 = \pm c p^a_0 = \hbar 
\omega_0$ is fulfilled, with the momentum $p^a_0$ satisfying the 
equalities $p^a_0 \equiv 
p^1\vert_{a = 0} = m_0 \dot{a} (\tau)\vert_{a = 0} = \pm m_0 c$,
respectively, and with 
the quantity $\omega_0$ being the circular frequency of the particle's 
oscillatory motion in the additional spatial dimension, which is 
measured in the particle's own (rest), or instantaneous
local inertial rest reference frame 
projected orthogonally onto the spacetime $a = 0$. On the other hand, 
it is worth adding that the above considerations allow one to define 
the ``dressed'', or renormalized rest mass $m_0$ of a particle as the
quotient $E_0/c^2$ of the ``internal'' 
(or rather directly unobservable) kinetic 
energy $E_0$ of the particle's {\it Zitterbewegung} divided by the 
squared velocity $\pm c$ of the {\it Zitterbewegung}; the particle's rest 
mass $m_0$ can equivalently be determined as the quotient $p^a_0/(\pm c)$ 
of the momentum $p^a_0$ of the particle's {\it Zitterbewegung} 
divided by the {\it Zitterbewegung} velocity $\pm c$; see also 
Ref.~\cite{Puthoff1}. Note that at the end of
section 7.6 we will present a somewhat different interpretation 
of the effect of {\it Zitterbewegung}.

Another interesting question concerns the possible tests of the 
toy model proposed in this paper; in order to handle it --
at least partially -- let us consider the spacetime with the 
metric (\ref{schw}). In the weak-field approximation, for which
one assumes that the inequalities $2 M/r \ll 1$ and 
$|\lambda| a^2 \ll 1$ are satisfied, the potential function
can be conventionally defined as follows,
\begin{equation}
V ( {\bf r} ) = - c^2 \frac{M}{r} + \frac{c^2}{2} |\lambda|
a^2 (\tau) \ ,
\label{potent}
\end{equation}
where $r \equiv | {\bf r} |$, the quantity ${\bf r} \equiv 
[x - x_0, y - y_0, z - z_0]$ denotes the three-dimensional position 
vector with respect to the centre of the mass $M$, which is situated 
at the point $(x_0, y_0, z_0)$, and the function $a = a (\tau)$ 
signifies the coordinate of the additional spatial dimension for
the particle(s) under consideration; it is worth mentioning here 
that one has $|d \tau| = (1 - 2 M/r)^{1/2} |d t| \cong (1 - M/r) |d t| 
\approx |d t|$ for $a = 0$ and $r, \theta, \varphi = {\it const}$.
Taking into account the results of sections 7.1 and 7.2 of this 
paper we can expect the quantity $a^2 (\tau)$ to oscillate rapidly 
in time, so one will detect the average value of this quantity over 
the proper time $\tau$, which we designate as $\langle a^2 (\tau) 
\rangle_\tau$.
Note that for a macroscopic (solid) body taken as a whole one
should have $a (\tau) \approx 0$, since such an object consists
of a large number of particles, between which the gravitational
``interactions'' are negligible -- hence, the superposition of all
the oscillations of object's particles is close to zero, especially
if we take into account all the intrinsic interactions within this
body, which contribute to its total energy $E$ and, in consequence,
are expected to decrease considerably the amplitude of the object's 
oscillatory motion in the additional spatial dimension. Thus, the use 
of the weak-field approach $|\lambda| a^2 \ll 1$ seems to be fully 
justified in the case of the investigated problem.

The gravitational-acceleration function ${\bf g} = {\bf g} ({\bf r})$
outside the mass $M$ can then be easily calculated and it reads
\begin{eqnarray}
{\bf g} ( {\bf r} ) = - \nabla_{\bf r} V ({\bf r}) \!\!\! &=&
\!\!\! - c^2 \frac{M {\bf r}}{r^3} - \frac{c^2}{2} \,
\nabla_{\bf r} \!\left[ \left\langle |\lambda| a^2 (\tau) 
\right\rangle_\tau
\! ({\bf r}) \right] \nonumber\\  \!\!\! &\equiv & \!\!\! 
- c^2 \frac{M {\bf r}}{r^3} - {\bf g}_l ({\bf r}) \ .
\label{accel}
\end{eqnarray}
Obviously, it is interesting to estimate the range of values and
of directions which can be assumed by the second term occurring 
in expression (\ref{accel}), i.e., by the function ${\bf g}_l = 
{\bf g}_l ({\bf r})$. It is clear that this term corresponds to local
interactions between particles which remain
outside the mass $M$ in the case being considered and whose behaviour 
is characterized, among others, by parameters such as the energies $U$ 
and $E$ of particles and by geometric properties of the investigated
system. On the one hand, the value of the function
${\bf g}_l ({\bf r})$ is, on average, expected to be equal to zero 
within an isotropic and homogeneous object which is isolated from all 
the long-range forces and remains in a state of
thermal equilibrium. On the other hand, it seems that the function 
$|{\bf g}_l ({\bf r})|$ can assume relatively large values in small 
regions of space, which are filled with highly anisotropic and/or
inhomogeneous, possibly different materials. Such a situation may 
occur, e.g., around the boundary between two or more samples of 
materials which remain in different thermodynamic phases, or in 
different density states; for instance, it may happen around the 
place of contact of a dense superconductor, or of a Bose--Einstein 
condensate with a hot low-density material remaining in an incoherent 
state -- for such cases we expect that the gradient of the quantity 
$\langle a^2 (\tau) \rangle_\tau ( {\bf r} )$ could take especially
large absolute values. 

In turn, it is easy to see that if we assume constant lengths for each 
of the two vectors entering expression (\ref{accel}) separately, then 
the acceleration $|{\bf g} ({\bf r})|$ should reach its maximum or 
its minimum value when the vectors $c^2 M {\bf r}/r^3$ and ${\bf g}_l 
({\bf r})$ are parallel or antiparallel to each other, respectively; 
obviously, it occurs when both the samples are situated in the same 
straight line crossing the centre $(x_0, y_0, z_0)$ of the mass $M$, 
so one of the objects (e.g., a coherent sample) is simply above the 
other with respect to the mass $M$, and the boundary between both the 
objects is perpendicular to the position vector $\bf r$. Then, it is 
important to add that according to expression (\ref{accel}) the 
gravitational field is expected to be modified -- either weakened or 
intensified -- also in some, possibly relatively large regions of 
space, which in the discussed case are situated mainly {\it above} 
and/or {\it below} the place of contact of both the samples. 

In general, one supposes as well that the presence 
of any external fields can significantly influence the value of the 
function ${\bf g}_l ({\bf r})$; for instance, a high-frequency 
electromagnetic field is expected to supply the energy (almost) 
homogeneously to the whole volume of the samples and would cause 
the occurrence of intense electric super-currents as well
as of other (e.g., collective) phenomena within the superconductor 
sample, which in turn could considerably change the values of the 
quantities contributing to the acceleration ${\bf g}_l ({\bf r})$.

In the context of the last few paragraphs, it is worth mentioning
that one can investigate as well the full form of the potential 
function in the weak-field approximation,
\begin{equation}
V ( {\bf r} ) = - c^2 \frac{M}{r} + \frac{c^2}{2} |\lambda|
a^2 (\tau) - c^2 \frac{M}{r} |\lambda| a^2 (\tau) \ ;
\label{pot_full}
\end{equation}
in such a case the (detectable) gravitational acceleration $\bf g$ 
outside the mass $M$ is given by the formula
\begin{eqnarray}
{\bf g} ( {\bf r} ) = - \nabla_{\bf r} V ({\bf r}) =
- c^2 \frac{M {\bf r}}{r^3} \!\left[ 1 + \left\langle |\lambda| 
a^2 (\tau) \right\rangle_\tau \! ({\bf r}) \right] - 
\left( 1 - \frac{2 M}{r} \right) \! {\bf g}_l ({\bf r}) 
\label{ac_full}
\end{eqnarray}
where the function ${\bf g}_l ({\bf r})$ has been defined 
in expression (\ref{accel}). 

Perhaps, the occurrence of non-vanishing values of the local
component ${\bf g}_l ({\bf r})$ of the gravitational acceleration
can be confirmed by experiment and is responsible for the existence
of the so-called weak shielding against the gravitational force, or 
the anomalous weight-loss effect which has been observed recently; see 
Refs.~\cite{Podklet,Li,Modan} and references therein, but notice
Ref.~\cite{Witteborn} as well. The phenomenon
described here may also be -- at least partially -- the reason for the 
problem with an accurate measurement of the Newtonian gravitational
constant $G$ as well as it may cause the occurrence of the effects
which are attributed to the existence of the so-called fifth force;
however, from the formula (\ref{ac_full}) it clearly results that the 
above-mentioned and/or similar (though very weak) effects can occur
also in the case when the condition ${\bf g}_l ({\bf r}) = {\bf 0}$ 
is satisfied. Note as well that the investigation of the potential 
function (\ref{potent}), performed in the context of the virial
theorem, allows one to explain the phenomenon of the so-called dark 
matter as arising due to the geometric properties of the spacetime 
which contains an additional spatial dimension; see Ref.~\cite{wt} 
for details.

\newpage
\noindent {\it 7.6.\ A model of an elementary particle}
\bigskip

Now we will try to obtain a na\"{\i}ve form of the particle's 
stability conditions mentioned in the previous subsection. To this 
end, first of all we recall from section 7.4 that an 
electromagnetic wave can most probably
be assigned to each of the two components, of which a photon 
seems to consist. However, when those two modes of a photon 
become two real, separate particles described by expressions
(\ref{par1}) and (\ref{par2}), then each of those two particles 
separately does not seem to be associated with any freely propagating
electromagnetic wave. Thus, in the case of a massive particle
we can expect that the trajectory of its accompanying 
electromagnetic wave is curved
in the gravitational ``field'' of that particle to such an extent
that it forms a circular orbit around the particle. In the case of 
a non-rotating charged particle, the circular orbits corresponding
to the condition $r = {\it const}$ are given by the stationary points 
of the Reissner--Nordstr\"om potential (for $a = 0$) \cite{MTW} which 
characterizes the spacetime where the electromagnetic wave is moving;
we thus demand the following condition to be fulfilled,
\begin{equation}
0 = \frac{d}{dr} \!\left[ \frac{1}{r^2} \!\left( 1 - 
\frac{2 M}{r} + \frac{Q^2}{r^2} \right) \right] \ ,
\label{RN}
\end{equation}
where the mass $M \equiv G m_0/c^2$ and the particle's (specific)
electric charge $Q$ are expressed in geometrized units. Note that 
in such units the expression for the fine-structure constant $\alpha$ 
takes the form
\begin{equation}
\alpha = \frac{e^2 c^3}{G \hbar}
\label{fine1}
\end{equation}
where the quantity $e$ signifies the observed elementary electric 
charge expressed in meters; the specific electric charge $e$ is related
to the electric charge $e_{\rm SI}$ expressed in SI units by the 
formula $e \equiv e_{\rm SI} [G/(4 \pi \epsilon_0 c^4)]^{1/2}$ where
$\epsilon_0$ denotes the electric constant (the permittivity of the 
vacuum). Obviously, equation (\ref{RN}) has the two solutions,
\begin{equation}
r_{\pm} = \frac{1}{2} \!\left( 3 M \pm \sqrt{9 M^2 - 8 Q^2}\,
\right) \ ,
\label{RNsol}
\end{equation}
the first from which is unstable, whereas the second one remains 
stable, respectively; note that we assume in this paper that the
existence of a {\it stable} circular orbit of an electromagnetic wave 
around the particle being considered means that such a particle 
satisfies the ``weak'' stability condition mentioned in section 7.5.

In order to maintain the ``strong'' stability of the orbit on
which the electromagnetic wave is moving around the charged particle,
we impose the following condition on the wavelength 
$\lambda_0$:\ $2 \pi r \, n = \lambda_0$ where $n \in {\bf N}$.  
Such a condition ensures that the length $\lambda_0$
of the electromagnetic wave is an integer multiple of the length 
$2 \pi r$ of the circular orbit, so the electromagnetic wave
forms a kind of a standing, or a ``frozen'' wave around the particle; 
thus, the particle can be expected to survive for some finite time, so 
also it is then supposed to be subject to observation. On the other
hand, let us consider the circular frequency $\omega = (c^2 p^2 +
m_0^2 c^4)^{1/2}/\hbar$ of the oscillatory motion of the 
particle with respect to the additional spatial dimension $a$ of 
the spacetime described, for instance, by the metrics (\ref{sol}) or 
(\ref{schw}) and assume that the quantity $\omega_0$ signifies the value 
of the circular frequency, which is measured in the particle's own
(rest), or instantaneous local inertial rest reference frame projected 
orthogonally onto the spacetime $a = 0$; such a reference system can 
symbolically be denoted as $[t_0 (\tau), a_0 (\tau), x_0 (\tau), 
y_0 (\tau), z_0 (\tau)] (a = 0)$ where the quantity $\tau$ stands 
for the affine parameter, or the proper time of the particle. From 
expressions (\ref{par1}) and/or (\ref{par2}) in section 7.5
we know that the circular frequency $\omega$ satisfies 
the equality $\hbar \omega_0 = m_0 c^2$,
since one has by definition that $p \equiv (- \sum_{i = 2}^4 g_{ii}
p^i p^i)^{1/2} = 0$ for $\omega = \omega_0$; 
in turn, from section 7.4 it is clear that the circular frequency 
${\widetilde{\omega}}$ of the electromagnetic wave moving around 
the particle fulfils the relationship ${\widetilde{\omega}}_0 = 
c {\widetilde{p}}_0/\hbar = 2 \pi c/\lambda_0$
where ${\widetilde{p}}_0$ denotes the value of the momentum
${\widetilde{p}}$ for ${\widetilde{\omega}} = {\widetilde{\omega}}_0$.
Let us assume that 
$\omega_0 = {\widetilde{\omega}}_0$, which means that the mass
$m_0$ of the particle being considered comes exclusively from 
the energy $\hbar {\widetilde{\omega}}_0$
of the electromagnetic wave moving on the circular orbit around the 
particle; see also the discussion contained in the last three
paragraphs of this subsection. Combining all the 
above requirements regarding the quantities $\lambda_0$, $\omega_0$,
${\widetilde{\omega}}_0$ and $m_0$ together with expression 
(\ref{RNsol}), one easily obtains the stability condition of 
the following form:
\begin{equation}
\frac{2 G \hbar}{3 M_n^2 c^3 n} = 1 \pm 
\sqrt{1 - \frac{8 Q^2}{9 M^2_n}} \, \ .
\label{stab}
\end{equation}
We expect that the mass $M_n$ in the above formula corresponds
to the ``bare'' masses of elementary particles, so it may change
with the index $n$, whereas the absolute value of the electric 
charge $Q$ should remain constant
for various elementary objects. In the case of $n = 1$, let us 
then identify the mass $M_1$ in Eq.~(\ref{stab}) with the greatest
mass possible in the toy model, which is the Planck (cutoff) one, 
$M_{\it Pl} \equiv E_{\it Pl}/c^2 = E_c/c^2$, so $M_{\it Pl} = 
(G\hbar/c^3)^{1/2}$ in geometrized units. Note, however, that in 
the model under consideration the ``bare'' mass $M$ of a particle 
should fulfil the conditions $2 \sqrt{2}\, |Q|/3 \le M \le M_{\it Pl}$;
it means that Eq.~(\ref{stab}) possesses solutions only for $n = 1$ (if 
we assume that $|Q| = {\it const}$), so the ``bare'' mass of an 
elementary particle should be equal to the Planck mass; see also 
Ref.~\cite{Puthoff1}, and especially footnote~24 therein. In such 
a case, it is easy to see that Eq.~(\ref{stab}) has 
a solution only for the stable circular 
orbit of the electromagnetic wave, whose radius is given by $r_{-}$
in the formula (\ref{RNsol}). Thus, we obtain the value of 
the particle's ``Planck electric charge'',
\begin{equation}
Q = \pm \left( \frac{G \hbar}{c^3} \right)^{\! 1/2} \ ,
\label{charge}
\end{equation}
which remains one order of magnitude greater than the value of
the observed elementary electric charge $e$,
\begin{equation}
Q = \frac{e}{\alpha^{1/2}} \ .
\label{obs_e}
\end{equation}
It is clear that the electric charge $Q$ actually corresponds to the 
``bare'' elementary electric charge, i.e., to the electric charge which 
is in principle expected to assume exactly the observed value $e$ of the 
elementary electric charge when ``dressed'' by the screening of 
the vacuum fluctuations, that is to say, when situated in a 
sufficiently large distance from an observer. Similarly, the ``bare''
mass of an elementary particle should be equal exactly to the Planck
mass $M_{\it Pl}$. It is important to note that for $n = 1$ the 
circular orbit of the electromagnetic wave would lie precisely on 
the event horizon of a Reissner--Nordstr\"om black hole, which is
given by the equation $(1 - 2 M/r + Q^2/r^2) = 0$; thus,
the particle being considered would form a micro-black hole 
and be indistinguishable from other particles of the same kind.

Of course, the above-described simple model of an elementary particle
does not take into account some important features of real particles,
such as an intrinsic angular momentum (spin); 
for instance, it seems that the latter quantity could be 
incorporated into the model by considering the Kerr 
metric for a rotating and uncharged black hole, assuming that such
an object represents an elementary particle. In such a case, it can 
be proved that the angular momentum $J$ of a ``bare'' particle 
surrounded in its equatorial plane by the stable circular orbit of 
the electromagnetic wave should fulfil the Regge-like relation 
$J = \pm M^2$ where $M = M_{\it Pl}$,
so one has $J_{\rm SI} = \pm G m_{\it Pl}^2/c = \pm \hbar$ with the 
angular momentum $J_{\rm SI}$ and the Planck 
mass $m_{\it Pl}$ expressed in SI units. We expect that 
such a value of the angular momentum $J$ denotes here the ``bare'' 
value of the particle's spin. It is clear that the circular orbit  
in the equatorial plane given by $\theta = \pi/2$ is the only one 
which is stationary in the case of the Kerr metric, since only for 
such an orbit the condition $p^\theta = 0$ holds always. If the 
relationship $J = \pm M^2$ is satisfied, then the circular orbit 
of the ``frozen'' electromagnetic wave in the equatorial
plane of the Kerr spacetime is given by the radius $r = M$ which 
corresponds in such a case to the event horizon of an extreme Kerr 
black hole; for such a situation both the inner and the outer event 
horizons of a Kerr black hole become identical with each other.
It is easy to conclude that the investigated
model of an elementary particle represented by an
extreme Kerr black hole implies that uncharged particles with 
a non-zero spin cannot be massless; the above statement concerns, 
for instance, neutrinos provided that they are described by 
the discussed black-hole model. Note also as a curiosity that the 
relationship $J = \beta M^2$ recovers at the microscale the Wesson 
data originally obtained for various astronomical objects of extremely 
different sizes \cite{Wesson}; here $\beta$ is a constant quantity 
which fulfils the condition $G/(\beta c) \sim \alpha \cong 1/137$.

Similarly, in the case of a rotating and charged particle, one
can consider the stationary axially-symmetric charged rotating 
Kerr--Newman solution of the Einstein equations (\ref{1}) to 
obtain the ``extreme'' relationship between the mass $M$, 
the angular momentum $J$ and the electric charge $Q$ of 
a ``bare'' elementary particle surrounded in its equatorial plane
by the stable circular orbit of the electromagnetic wave; this
relationship reads $M^2 = (J/M)^2 + Q^2$ where $M = M_{\it Pl}$,
so it can then be written as
\begin{equation}
1 = \left( \frac{J_{\rm SI}}{\hbar}\right)^{\! 2} + 
\frac{Q_{\rm SI}^2}{4 \pi \epsilon_0 \hbar c} \ ;
\label{KNsol}
\end{equation}
it is interesting to note that the Newtonian gravitational 
constant $G$ does not enter the above equation. 
It should also be remarked that, similarly as in the case of 
the Kerr solution described above, the circular orbit of the 
``frozen'' electromagnetic wave in the equatorial plane of the
Kerr--Newman spacetime is given by the 
radius $r = M$ which corresponds in the case of the Kerr--Newman 
metric to the event horizon of an extreme Kerr--Newman black 
hole; for such a situation both the inner and the outer event 
horizons of a Kerr--Newman black hole become identical with each other.

Note that for a Kerr--Newman black hole with a mass $M$,
an electric charge $Q$, a magnetic dipole moment $\cal M$ 
and an angular momentum $J$, the following relationship is satisfied,
\begin{equation}
\frac{\cal M}{J} = \frac{Q}{M} \ ;
\label{gyro}
\end{equation}
it means that a Kerr--Newman black hole has the gyromagnetic 
ratio equal to $2$, just as an electron.  
In general, it is known that the Kerr--Newman metric describes 
properly the gravitational and electromagnetic fields of an 
electron, including the anomalous gyromagnetic ratio; see
Refs.~\cite{Carter,MTW} and also Ref.~\cite{LB} with references
therein. Note as a curiosity that the relationship
${\cal M}_{\rm SI}/J_{\rm SI} = Q_{\rm SI}/M_{\rm SI} \sim
G^{1/2}$ obtained for the ``bare'' values of the above four
quantities recovers, to within a proportionality 
constant, the so-called Blackett--Sirag
relation $({\cal M}_{\rm SI}/J_{\rm SI})_{\it astro} \sim
G^{1/2}$ which holds for the Blackett and Wesson data
concerning the gyromagnetic coefficient of various
astronomical objects; see Ref.~\cite{Sirag}. On the other 
hand, the relationship $J \propto M^2$ corresponds to the 
so-called Regge trajectories of hadrons, which fact seems to 
be quite relevant in regard to the subject of this subsection.
Note also that there exists a gravitational ``spin--spin'' force
which is attractive for antiparallel spins \cite{Wald},
so the particle and its antiparticle in our model should
attract each other with this interaction as well, and not only
with the electromagnetic and the ``usual gravitational'' 
forces; of course, the final product of an annihilation --
photon(s) -- should then be scalar particle(s). It is worth 
adding as well that the most fundamental particles 
-- i.e., the electron, the positron and the neutrino/antineutrino 
-- can be used to generate the mass spectrum of all elementary 
particles; see Ref.~\cite{Berg}. One can also conclude easily
that there do not exist particles with the properties of
magnetic monopoles within the here-proposed toy model of elementary
particles, since there are no black-hole solutions of Eqs.~(\ref{1})
which allow for the existence of single and separate (isolated) 
``magnetic charges''. Similarly, one does not expect to incorporate 
uncharged elementary particles without spin into the model being 
considered, since the only closed (circular) orbit of the 
electromagnetic wave is unstable in the case of the Schwarzschild 
metric. And indeed, it is interesting to note that the uncharged 
spinless elementary particles such as the mesons $\pi^0$, $K^0$, or 
$\eta^0$ have lifetimes significantly shorter than most of the unstable
elementary particles, each of which possesses a non-zero spin and/or 
a non-vanishing electric charge; it is clear that one expects at least 
the ``weak'' stability condition to be satisfied in the case of 
the latter particles.

We suppose that the mass (or the total rest energy $E_m$) of 
the investigated black hole representing a toy-model stable particle
should remain constant independently
of the possible changes of the black-hole thermodynamic
parameters, such as the area of its event horizon. Thus, 
the specific heat of the black hole should be equal to zero, which 
occurs in the case of Kerr--Newman black holes when the condition
$M^2 = (J/M)^2 + Q^2$ is fulfilled, i.e., exactly in all the 
cases considered in this paper; note that for such a condition
the thermodynamic temperature, or the surface gravity $\cal K$
of a black hole is equal to zero; see Ref.~\cite{Wald}. 
It is worth adding that the thermodynamic temperature of
particles represented by non-extreme micro-black holes would
be finite (non-zero) \cite{Wald}, so such particles are expected to
evaporate due to the Hawking effect \cite{Hawking}; the occurrence
of the above-mentioned evaporation process in the case of non-extreme 
black-hole particles indicates that they are unstable (like
resonances and some other unstable particles), which is consistent 
with the fact that they do not fulfil the stability condition(s) for 
the circular orbit of the ``frozen'' electromagnetic wave in the 
particle's equatorial plane. We can also notice that the 
information content of the surface (i.e., of the event horizon) 
of a black hole of the Planck size is equal exactly 
to one classical $c$-bit of the Shannon--Bekenstein information.

It would be interesting to calculate the values of the 
quantities $J$ and $Q$ entering Eq.~(\ref{KNsol}). We can
try to achieve this by estimating the contribution of
each of those two quantities separately into the mass 
$M = M_{\it Pl}$ of the black hole being considered. To this end, 
we will employ the concept of the so-called irreducible mass 
$M_{\it ir}$ which is defined by the equation
\begin{equation}
M^2 = \left( M_{\it ir} + 
\frac{Q^2}{4 M_{\it ir}}\right)^{\! 2} + 
\frac{J^2}{4 M^2_{\it ir}} \ ;
\label{irr}
\end{equation}
see Refs.~\cite{Chris1,Chris2,MTW}. Thus, the contribution $M_Q$
of the electric charge $Q$ to the mass $M = M_{\it Pl}$, when 
assuming the conditions $Q^2 = M^2$ and $J = 0$ to be fulfilled, is 
equal to $M_Q \equiv M - M_{\it ir} = M/2$. Similarly, the 
contribution $M_J$ of the angular momentum $J$ to the
black-hole mass $M$, under the assumptions that $J^2 = M^4$ and
$Q = 0$, can easily be calculated to be equal to $M_J \equiv
M - M_{\it ir} = M (1 - 2^{-1/2})$. Performing the appropriate
normalization procedure resulting from Eq.~(\ref{KNsol}), we then
arrive at the following values for the ``bare'' fine-structure
constant $\alpha_{\rm SI}$ and the ``bare'' intrinsic angular
momentum $J_{\rm SI}$, both ones being expressed again in SI units,
\begin{eqnarray}
\alpha_{\rm SI} \!\!\! & = & \!\!\! \frac{Q_{\rm SI}^2}{4 \pi 
\epsilon_0 \hbar c} = \frac{1}{7 - 4 \sqrt{2}} \cong 0.74452
\label{fine2} \\
J_{\rm SI} \!\!\! & = & \!\!\! \pm \frac{2 - \sqrt{2}}{\left(7 - 
4 \sqrt{2}\,\right)^{\! 1/2}} \, \hbar \cong 
\pm 0.50545 \, \hbar \ ,
\label{spin}
\end{eqnarray}
respectively. Note that equation (\ref{KNsol}) together with
the above-described considerations resulting in expressions
(\ref{fine2}) and (\ref{spin}) seem to explain the apparently
random coincidence which manifests itself in assuming the discrete 
values actually by {\it both} the particle's electric charge $Q$ 
and spin $J$. 

It should now be remarked that the increase in
the value of the fine-structure constant $\alpha$ with
rising value of the interaction energy (or the momentum transfer)
has been confirmed by experiment, at least for the electroweak
interactions; see Ref.~\cite{Levine}. It would then be interesting 
to compare the result given by the formula (\ref{fine2}) with
an analytic extension of the results obtained while employing
the method described in Ref.~\cite{Levine}, for experiments
corresponding to energies as high as it is possible. Similarly,
the ``bare'' value of the angular momentum $J$ of an elementary
particle determined in expression (\ref{spin}) can be verified
by high-energy measurements of the value of a single spin; some
(limited) kinds of such experiments are currently planned to
be performed; see, for instance, Ref.~\cite{Kane} and references
therein. However, a somewhat more promising approach to verifying
the theory of an elementary particle presented in this section
would consist in experimental testing the relation (\ref{gyro}),
with simultaneous (but possibly separate) measurements of all the 
factors ${\cal M}, J, Q, M$ entering that formula, for as wide 
as possible range of the interaction energy (the momentum transfer) 
of an elementary particle, e.g., of an electron. It should be taken 
into account here the fact that a boost (in the spin direction) with 
the velocity $v$ transforms the Kerr--Newman parameters $J/M$ and $M$ 
according to the formulae $(J/M)' = (J/M) (1 - v^2/c^2)^{1/2}$ and 
$M' = M (1 - v^2/c^2)^{-1/2}$, respectively; see 
Refs.~\cite{Burin,Arcos}.

It would also be interesting to develop the ideas of the geometric 
model of an elementary electric charge, presented in 
Refs.~\cite{JAW1,JAW2}, in the context of the toy model being considered 
whose topology incorporates an additional spatial dimension. Note that 
the concept of ``charge without charge'' discussed in Ref.~\cite{JAW1} 
was partially employed in a classical model of an electron proposed 
in Ref.~\cite{Arcos} where the electron's electric charge is associated
with the net flux of an electric field which is topologically trapped 
in the naked circular (ring) singularity of the maximally extended 
Kerr--Newman spacetime \cite{Wald}. Of course, a similar 
interpretation of the electric charge $Q$ can also be applied in the 
toy model presented in this paper; cf.\ section 7.8.
Perhaps, the existence of a topological connection -- like those 
described in Refs.~\cite{JAW1,Arcos} -- between particles which have 
been previously interacting, could serve as a basis for a possible 
explanation of the Einstein--Podolsky--Rosen paradox \cite{EPR}, 
at least from a purely theoretical point of view.

The concept of ``mass without mass'' presented in Ref.~\cite{JAW1}
encourages us to stress that the particle's mass, or total (rest) 
energy $E_m$ in the toy model can be interpreted as consisting merely 
of the gravitational, electromagnetic and rotational energies, 
each of which is confined to the region of space bounded in the 
particle's equatorial plane by the circular orbit of the ``frozen'' 
electromagnetic wave, which orbit coincides with the event horizon 
of an extreme Kerr--Newman black hole of the Planck size; it seems 
that the total rest 
energies of each particle and of its antiparticle can come from the 
contributions of the quantity $U_{1 2}$, as discussed in section 7.5. 
In the context of this paragraph, it is reasonable to mention 
interesting considerations concerning the concept and the possible 
electromagnetic origin of mass and inertia, which are contained in 
Refs.~\cite{Puthoff2,Haisch1,Haisch2} and in references therein. One 
should also note that several advanced models of an electron and of
other elementary particles, based on the Kerr--Newman topology,
have already been proposed in the literature; 
see, for instance, Ref.~\cite{Sid} as well as 
Ref.~\cite{Arcos} and references therein. Another interesting models
of extended elementary particles based, among others, on the 
generalized classical electrodynamics, on the group theory,
on various topological structures, on wave mechanics,
or on the Dirac--Maxwell field formalism are 
presented in Refs.~\cite{Gallop,Roman,Jehle,Cornil1,Bohun},
respectively; see also some of numerous references therein.
In the context of this paper, the particularly relevant 
references \cite{Ashw,Jennis1,Jennis2} are worth mentioning as well.
One can also recall an interesting model in which the zero-point
vacuum fluctuations are considered as a possible source of the 
acceleration of polarizable cosmic-ray particles, such as protons; 
see Ref.~\cite{Rueda}.

Here we return for a moment to the effect of the {\it Zitterbewegung}
of a massive particle, which has been briefly analyzed in section 7.5. 
It is well-known that the velocity eigenvalues in the
Dirac equation are determined to be equal to $\pm c$. This fact may
simply correspond in the toy model here-presented to the 
existence of the ``frozen'' electromagnetic wave which is curved
in the gravitational ``field'' of a particle to form the circular
orbit in the particle's equatorial plane. Obviously, in such a case
the notion of {\it Zitterbewegung} would be used in a completely
misleading way to call a phenomenon that is entirely different
from the one for which this term was intended; a somewhat better
name for the effect associated with the curved electromagnetic
wave would be, for instance, an {\it Umkreisbewegung}.

\bigskip
\bigskip
\noindent {\it 7.7.\ Thermodynamic properties of particles}
\bigskip

Let us consider a free test particle which is moving with
a uniform rectilinear motion in regard to the spacetime
$a = 0$ that is embedded in the five-dimensional manifold
with the metric (\ref{sol}). According to the formula (\ref{mot1})
which is a solution of the integrated equation of motion (\ref{enr}),
such a particle simultaneously oscillates in the additional spatial
dimension $a$. Obviously, the oscillatory motion of a particle is 
characterized, among others, by a proper acceleration which is defined 
as ${\ddot{a}} (\tau) \equiv d^2 a(\tau)/d\tau^2$. Due to the rapid
oscillations of the sine function entering the formula (\ref{mot1}) 
in the proper time $\tau$, or with respect to an affine parameter 
in general (including the one for massless particles), it seems 
reasonable to assume that it is the average absolute value of 
the proper acceleration (or deceleration) over the affine parameter 
$\tau$ rather than the ``pure'' absolute value $|{\ddot{a}} (\tau)|$
of the proper acceleration, what should 
be regarded as a quantity experienced by a free particle for 
sufficiently long periods of the proper time, or for sufficiently large
intervals of the affine parameter. The average absolute 
value of the particle's proper acceleration is easy to calculate
and it reads
\begin{equation}
\chi_{\it av} (U, E) \equiv \left\langle \left\vert {\ddot{a}} 
(\tau) \right\vert \right\rangle_{\tau} = \frac{2 E}{\pi \hbar^2} 
\sqrt{\frac{U^2 - E^2}{|\lambda|}} \, \ .
\label{avac}
\end{equation}
It is then reasonable to recall now the Unruh--Davies effect which 
predicts that a uniformly accelerated test particle moving in the 
ordinary Minkowski vacuum is expected to find itself immersed in a 
bath of the thermal radiation with an absolute temperature proportional
to the particle's proper acceleration relative to a Lorentz frame; 
see Refs.~\cite{Davies1,Unruh1,Soffel,Don}. One should note that 
the Unruh--Davies effect seems to be of a fundamental rather than
of an artificial nature; for instance, it is not limited to a 
free-field theory \cite{Wichm} and has been generalized to a
curved spacetime \cite{Wald1,Wald2}. It is also worth adding
that there seems to exist a classical counterpart of the 
Unruh--Davies phenomenon in electrodynamics; see Ref.~\cite{Lin}.
In this subsection we will try to apply the predictions resulting
from the Unruh--Davies effect to the case of the oscillatory
motion of a free particle which is travelling in the spacetime 
described by the metric (\ref{sol}). Thus, instead of considering the 
geodesic motion of a particle in the curved spacetime, we rather
assume here the point of view -- remaining in accord with the 
principle of equivalence -- that the investigated particle  
is moving in the ordinary Minkowski vacuum with the average proper 
acceleration given by the formula (\ref{avac}). 

First of all, we should note that the proper acceleration 
${\ddot{a}}_{1 2}$ of a {\it virtual} particle--antiparticle 
pair considered in sections 7.2 and 7.3 is equal to zero, since one 
has ${\ddot{a}}_{1 2} (\tau) \equiv {\ddot{a}}_1 (\tau) + 
{\ddot{a}}_2 (\tau) = 0$ for any value of the affine parameter $\tau$; 
obviously, in such a case the equality $\langle \vert \ddot{a}_{1 2} 
(\tau) \vert \rangle_\tau = 0$ holds as well. 
Thus, according to the Unruh--Davies effect 
extended to the toy model, the absolute temperature $T$ detected by a 
virtual particle--antiparticle pair remaining as a whole at rest in some 
vacuum inertial reference frame is equal to zero, since we expect that 
in such a case the relationship $T \propto |{\ddot{a}}_{1 2} (\tau)|$ is 
satisfied. The above simple remarks lead one to the self-consistent
conclusion that the temperature of the vacuum remains actually equal
to zero. In turn, let us now consider a large enough set, or a gas of 
non-interacting {\it real} (non-virtual)
particles which are in thermal contact
with one another, though initially may not be in a state of thermal
equilibrium. We suppose that after the lapse of some finite 
time, the set of particles -- whose parameters can change during 
the (elastic) collisions, or encounters occurring between them --
will remain in thermal equilibrium, at some temperature $T$,
with a thermal bath which arises due to the Unruh--Davies effect 
occurring for the particles under consideration; see, for instance, 
Ref.~\cite{Raine}. Then, depending on the type of the particles, the 
distribution of the energies $E$ of the 
particles is expected to be static with a bosonic-like \cite{Unruh1}
or with a fermionic-like \cite{Soffel} energy spectrum, whose densities
are given by the Unruh--Davies formulae
\begin{equation}
\rho_\chi (E) \equiv \left( \exp \!\left[ \frac{2 \pi c \left( 
E - E_0 \right)}{\hbar \chi} \right] \mp 1 \right)^{\! -1} \ ,
\label{acdist}
\end{equation}
respectively, where the quantity $\chi = \chi (U, E)$ here denotes 
the value of the uniform proper acceleration of a particle with the 
energy $E$ as well as with the hidden parameter $U$, and the 
symbol $E_0$ stands for the ground-state energy of each of 
the particles being considered; for instance, we expect that the 
relationship $E_0 = m_0 c^2$ holds in the case of a gas of elementary 
particles where the quantity $m_0$ is the rest mass of each particle
of the gas. The form of the functions $\rho_\chi (E)$ given by 
expression (\ref{acdist}) suggests strongly that they are closely 
related to the Bose--Einstein or the Fermi--Dirac
familiar energy distributions which read
\begin{equation}
\rho_T (E) \equiv \left[ \exp \!\left( \frac{E - \mu_c}{k_B T} 
\right) \mp 1 \right]^{\! -1} \ ,
\label{endist}
\end{equation}
respectively, where the quantity $k_B$ denotes the Boltzmann 
constant and the function $\mu_c = \mu_c (T)$ is the chemical,
or thermodynamic potential conventionally defined.

Let us assume that the gas being considered consists of $N$ particles
which occupy some volume $V$; the possible energy states of a single
particle are labelled with the use of the index $i$ and the energy
of a particle in a state $i$ is denoted by $E_i$. The thermodynamic
potential $\mu_c$ is then determined by the constraint on the total
number of particles, which can be written as $\sum_i \rho_T (E_i) 
= N$ where the above sum extends over all possible energy states
$i$ with the energies $E_i$ that satisfy the conditions $E_0 \le
E_i \le E_c$. We now assume that the energy distributions $\rho_\chi
(E)$ and $\rho_T (E)$ given, respectively, by expressions (\ref{acdist})
and (\ref{endist}) are equivalent to each other. Thus, one can easily
calculate the temperature $T$ of the thermal radiation
detected by a particle with the energy $E_i$, which is moving with
the uniform proper acceleration $\chi_i = \chi (U_i, E_i)$ in the 
special-relativistic four-dimensional spacetime that is given 
by $a = 0$ assumed in the metric (\ref{sol}); we obtain that
\begin{equation}
T = \frac{\hbar \chi_i}{2 \pi c k_B} 
\frac{E_i - \mu_c}{E_i - E_0} \ .
\label{temp}
\end{equation}
However, it is clear that the relationship (\ref{temp}) becomes 
merely approximate, if we assume that the quantity $\chi$ denotes 
therein the average absolute value $\chi_{\it av}$ of the 
particle's proper acceleration, given by the formula (\ref{avac}), 
rather than the (constant) value of a uniform proper acceleration. 

To proceed further, let us assume that the relationship 
$(U_i^2 - E_i^2) = A_i^2 E_{\it Pl}^2$ holds for each gas's particle 
which remains in an energy state denoted by the index $i$; 
according to the considerations concerning
the phenomenon of the {\it Zitterbewegung}, which are contained in
section 7.5, we expect that the values which can be taken by the 
function $A_i = A (U_i, E_i)$ should most probably satisfy for any $i$ 
the inequalities $0 < A_i \le 1$. Assuming that $\chi (U, E) \approx
\chi_{\it av} (U, E)$ and substituting the formula (\ref{avac}) 
into Eq.~(\ref{temp}), we then obtain easily that 
\begin{equation}
\left\langle E \right\rangle_E \approx \left\langle 
\frac{E - E_0}{A E} \right\rangle_{\! E} \pi^2 k_B T + \mu_c
\label{therm}
\end{equation}
where the mean value $\langle X \rangle_E$ of a quantity $X$ over
one of the two energy distributions $\rho_T (E)$ is given by
the formula
\begin{equation}
\langle X \rangle_E \equiv \frac{\sum_i X_i \, \rho_T \!\left( E_i
\right)}{\sum_i \rho_T \!\left( E_i \right)} \ ;
\label{aver}
\end{equation}
here the symbol $X_i$ denotes the value of the variable, or function
$X$ for an energy state $i$,
both the above sums range over all possible energy states $i$
and the summation limits are determined by the states with the 
energies $E_0$ and $E_c$, since the conditions $E_0 \le E_i \le E_c$ 
should be fulfilled for any energy state $i$. It is clear that for a
sufficiently large number of particles, the sums occurring in
expression (\ref{aver}) should be
replaced by appropriate integrals which contain the 
density-of-states function $g = g (E)$, exactly as in the case of 
the ordinary statistical thermodynamics; one would then have 
\begin{equation}
\left\langle X \right\rangle_E \equiv \frac{\int_{E_0}^{E_c} 
X \rho_T (E) \, g(E) \, dE}{\int_{E_0}^{E_c} \rho_T (E) 
\, g(E) \, dE} \ . 
\label{aver_int}
\end{equation}

Note that on the basis of Eq.~(\ref{therm})
we can suppose that the quantities $(E - E_0)/E$ and $A$ are 
represented by two distinct functions of temperature, 
which nevertheless behave in a similar way, at least in the 
high-temperature limit or for a sufficiently low concentration of the
gas being considered; in such cases one expects the quantity 
$\langle (E - E_0)/(A E) \rangle_E$ to be of the order of $10^{-1}$.
It is also worth noting that for the special case of a gas of photons 
we obviously do not require the total number of particles contained in 
some volume $V$ to be conserved, so each of the denominators occurring 
in expressions (\ref{aver}) or (\ref{aver_int}) can take any (positive)
value in such a case. Additionally, for a gas of photons one expects 
the equality $\mu_c (T) = 0$ to be fulfilled for any equilibrium 
temperature $T$ of the gas; according to the formula (\ref{therm}),
however, we obtain that $\mu_c (T) \approx (3 - \pi^2 \langle A^{-1}
\rangle_E ) k_B T$ for the perfect gas of massless (bosonic) particles 
which satisfies the caloric equation of state given by the expression
$\langle E \rangle_E = [\pi^4\zeta^{-1}(3)/30] k_B T \approx 3 k_B T$,
and derived with the use of the formula (\ref{aver_int}) for
$g (E) \propto E^2$, $E_0 = 0$. 
Consequently, one would have $0 < \langle A^{-1} \rangle_E < 1$ 
for a gas of photons, or even can suppose that $0 < A_i^{-1} < 1$ in 
the case of each gas's particle, which seems to be consistent with the 
assumption that the hidden energy $U$ of a photon contains the quantity
$U_{1 2}$; see sections 7.3--7.6.

It is easy to notice that equation (\ref{therm}) looks somewhat similar
to the formula resulting from the classical-thermodynamics principle
of equipartition of energy, which can be written as 
$\langle E_{\it kin} \rangle_B \propto k_B T$ 
where the kinetic energy $E_{\it kin}$ is defined as 
$E_{\it kin} \equiv E - E_0$, the value of the proportionality constant 
remains of the order of unity, and the 
symbol $\langle \cdot \rangle_B$ here denotes the mean value 
over the Boltzmann distribution of the energy $E$; for instance, 
in the case of a classical system remaining in 
thermal equilibrium at the absolute temperature $T$ which is much
lower than the Planck temperature defined as $E_c/k_B$, we easily
obtain that
\begin{eqnarray}
\left\langle E_{\it kin} \right\rangle_B \!\!\! &\equiv&
\!\!\! \frac{\int_{E_0}^{E_c} 
E_{\it kin} \exp \!\left[ - E/\!\left( k_B T \right) \right] 
g(E) \, dE}{\int_{E_0}^{E_c} \exp \!\left[ - E/\!\left( k_B T 
\right) \right] g(E) \, dE} \nonumber\\
\!\!\! &\cong& \!\!\! \left\{ \begin{array}{lcl}
k_B T & \;\; \quad \mbox{\rm for} & g (E) = 1 \\
k_B T/2 & \;\; \quad \mbox{\rm for} & g (E) = \sqrt{E_0/\!\left[2 
c^2 \!\left( E - E_0 \right)\right]} \, \ ;
\end{array} \right. 
\label{array}
\end{eqnarray}
here the former (approximate) equality represents the particle's mean 
energy for the energy $E$ which is randomly distributed among all the 
available energy states, whereas the latter formula determines the 
value of the mean kinetic energy per each independent degree of 
freedom, which is associated with any quadratic term occurring in the 
expression for the energy $E$ of the ideal gas of classical particles.
Taking into account the formula (\ref{therm}) one should then expect 
the relation $( \langle E \rangle_E - \mu_c ) 
\sim \langle E_{\it kin} \rangle_B$ to be satisfied, at least in 
the classical limit, i.e., for a sufficiently high temperature and/or 
for a low enough concentration of the gas; in such
cases we then have $\mu_c \sim E_0$. In turn, the actual value of 
the proportionality factor $\pi^2 \langle (E - E_0)/(A E) \rangle_E$ 
occurring in the formula (\ref{therm}) 
obviously depends on the type of the particles being considered 
as well as on the values of the parameters characterizing the particles, 
such as the energies $U$, $E$ and the chemical-potential function(s) 
$\mu_c$; each of the above-mentioned quantities is additionally 
expected to be a function of,
or to depend on the temperature $T$. The knowledge of the 
values of the quantities $U$, $E$ and of their general properties would
be particularly important in the context of an appropriate evaluation of 
the average proper acceleration of particles:\ note that the formula 
(\ref{avac}) is surely approximate only, especially if we take into 
account the fact that the estimation of the quantity $\chi$ has been 
performed with the use of the relation $\chi \approx \chi_{\it av}$; 
nevertheless, it allows one to obtain the relationship (\ref{therm}) 
which is indeed roughly consistent with the predictions of statistical 
thermodynamics.

On the other hand, the possible interactions between different 
particles within a set of particles would most probably exert an 
influence on the value of the energy $U$ of each single particle 
being subject to those interactions as well as -- maybe -- 
of the whole set of particles, since the quantity $U$ is 
a hidden variable, and/so the toy model seems to be a {\it non-local} 
theory (obviously, with respect to the spacetime $a = 0$ considered
alone, or separately). Clearly, we do not expect the formula 
(\ref{therm}) to be exact, as numerous simplifying assumptions 
have been made while deriving it. Some of the limitations of the 
above-investigated model are similar to those considered in 
Refs.~\cite{Dey1,Dey2} where a nucleon model has been proposed in 
which the partons (i.e., the quarks/antiquarks and gluons) are 
described as a gas, inside the confining hadron, remaining at finite 
temperature that arises due to the Unruh--Davies effect. 
Other limitations concerning the Unruh--Davies phenomenon, which can 
possibly influence the thermodynamic model described in this 
subsection, are also well-known; see, for instance, 
Refs.~\cite{Dey1,Dey2,Davies2,Rosu,Unruh2} and references therein. We 
should remember as well that the oscillatory motion of each particle
in the additional spatial dimension is perpendicular, or transverse
to the hypersurface $a = 0$, exactly as the vibrations of 
the vacuum fluctuations are; see section 7.2. Nevertheless, one should 
stress the promising agreement as to the order of magnitude of both 
the expected thermodynamic and the Unruh--Davies temperatures, 
the latter being detected and then assumed by the toy-model particles 
under consideration; such a 
convergence of predictions of the two evidently distinct theories 
occurs, among others, because of the enormous average absolute value 
of the proper acceleration which characterizes the oscillatory motion
of each toy-model particle in the additional spatial dimension. 

Thus, the considerations contained in this subsection lead us to the 
conjecture stating that the thermodynamic properties of a gas of 
toy-model particles can be a consequence of the oscillatory motion 
of each of such particles in the additional spatial dimension.
Note that results and conclusions analogical to those obtained in the
present subsection hold in the case of the spacetime described by the 
metric (\ref{schw}), with taking into account the fact that the energy
$E$ is given there by expression (\ref{schw_enr}), i.e., one has
$E = \hbar \omega (1 - 2 M/r)^{1/2}$ as well as that each of the 
quantities $E$, $E_0$ and $\mu_c$ occurring in the formulae 
(\ref{acdist}) and (\ref{endist}) should be divided by the factor 
$(1 - 2 M/r)^{1/2}$; for instance, one can easily derive equation
(\ref{therm}) where all the variables and/or parameters $A$, $E$, 
$E_0$ and $\mu_c$ are divided by the above factor.
Clearly, the subject of the investigation 
performed in this section needs further detailed analyses.

\bigskip
\bigskip
\noindent{\it 7.8.\ Electromagnetic phenomena and the vacuum
fluctuations}
\bigskip

According to section 3 as well as to subsections 7.2 and 7.3 of this 
paper, we know that the spacetime with the metric (\ref{sol}) is filled
with the virtual vibrations whose motions in regard to the 
additional spatial dimension are described by the formulae (\ref{mot1})
or (\ref{mot2}); here it is also worth adding that in the 
spacetime $a = 0$ and for constant values of the coordinates 
$x$, $y$ and $z$, the relationship $d \tau = d t$ is 
satisfied for both virtual and real particles as well as for
virtual antiparticles and the relation $d \tau = - d t$
holds for real antiparticles, where the symbol $\tau$ denotes 
the affine parameter. One then can note easily that the trajectory 
$a = a (\tau)$ of each of the virtual excitations being considered is 
actually equivalent to the trajectory, or rather to the values assumed 
by a homogeneous scalar plane wave $\psi = \psi (t, {\bf r})$ which 
propagates in the four-dimensional spacetime $a = 0$; such a wave 
can in some inertial reference frame of the spacetime $a = 0$ be 
denoted as $\psi \propto \sin \varphi \equiv \sin ( E t /\hbar 
\mp {\bf p} \cdot {\bf r}/\hbar )$ where the quantities $E$ and 
$\bf p$ are the energy and the momentum vector of an excitation, 
respectively, the symbol $\varphi = \varphi (t, {\bf r})$ 
signifies the phase function and the position vector $\bf r$ is
defined as usual as ${\bf r} \equiv [ x, y, z ]$. In turn,
the signs ``$\mp$'' in the above expression
refer to two opposite spatial directions of the motion of a virtual
particle or of its antiparticle. According 
to section 7.2 of this paper and to 
Ref.~\cite{wt}, the relationships $E = \pm \hbar \omega$ and 
${\bf p} = \pm \hbar {\bf k}$ are satisfied where the signs 
``$+$'' or ``$-$'' correspond to virtual particles or to their 
antiparticles, respectively (of course, the sign ``$+$'' in
the above formulae holds also for real particles as well as 
for real antiparticles). Thus, the wave being considered is 
characterized in some inertial reference frame of the spacetime 
$a = 0$ by a circular frequency $\omega$ and by a
wave vector $\bf k$ defined as ${\bf k} \equiv 2\pi 
{\bf n}/\lambda$ where the quantity $\lambda$ denotes the length
of the investigated wave and $\bf n$ is the unit vector oriented 
along the positive axis in the direction of the wave (or the 
excitation) normal. 

It is then clear
that the vibrations corresponding to a virtual particle or to 
its antiparticle, whose oscillatory motions with respect to the 
additional spatial dimension $a$ are described respectively
by the formulae (\ref{mot1}) or (\ref{mot2}), can be represented by
homogeneous scalar plane waves of the following forms:
\begin{eqnarray}
\psi^1 \!\left( t, \pm {\bf r} \right) \!\!\! &=& \!\!\! 
D \sin \!\left( \omega t \mp {\bf k} \cdot {\bf r} \right)
\label{wave1} \\
\psi^2 \!\left( t, \pm {\bf r} \right) \!\!\! &=& \!\!\! 
D \sin \!\left( - \omega t \pm {\bf k} \cdot {\bf r} \right) \ ,
\label{wave2}
\end{eqnarray}
respectively, where the quantity $D = D (U, E)$ denotes the 
wave-amplitude function. In turn, the signs ``$\mp$''
or ``$\pm$'' preceding the term ${\bf k} \cdot {\bf r}$ in the 
representations (\ref{wave1}) or (\ref{wave2}) correspond in each of
both the above cases to the two waves moving in opposite directions; 
note also that for the investigated vibrations one has 
$\psi^1 (t, {\bf r}) = \psi^2 (- t, - {\bf r})$. In the reference 
frame being considered, the superposition of the two waves moving in 
opposite directions forms a standing-wave pattern for each of the two 
wave-functions $\psi^1$ or $\psi^2$ separately; we then have
\begin{eqnarray}
\psi^1_s \!\left( t, {\bf r} \right) \!\!\! &\equiv& \!\!\! 
\psi^1 \!\left( t, {\bf r} \right) + 
\psi^1 \!\left( t, - {\bf r} \right) = 2 D \cos \!\left( {\bf k}
\cdot {\bf r} \right) \sin \!\left( \omega t \right)
\label{stand1} \\
\psi^2_s \!\left( t, {\bf r} \right) \!\!\! &\equiv& \!\!\! 
\psi^2 \!\left( t, {\bf r} \right) + 
\psi^2 \!\left( t, - {\bf r} \right) = 2 D \cos \!\left( {\bf k}
\cdot {\bf r} \right) \sin \!\left( - \omega t \right) \ ;
\label{stand2}
\end{eqnarray}
note that in another inertial reference frame of the spacetime $a = 0$
each of the above standing waves will move with a constant 
velocity, or with a uniform rectilinear motion, i.e., will have an
appropriately Lorentz-transformed four-vector of the
spacetime coordinates $c t$ and $\bf r$. Obviously, in the case 
of a pair of a virtual particle and its antiparticle one has $\psi^1_s 
(t, {\bf r}) + \psi^2_s (t, {\bf r}) = 0$ for any point $(t, 
{\bf r})$ of the four-dimensional spacetime $a = 0$, which 
corresponds clearly to the equality $a_{1 2} (\tau) = 0$ that
is discussed in section 7.2 of this paper. According to section 7.2,
we also expect that the wave-functions $\psi$ representing different 
pairs of virtual excitations (i.e., differing in at least one of 
the following three quantities:\ the value of the circular frequency 
$\omega$ as well as the length and the direction of the wave vector 
$\bf k$) can differ in values of the wave amplitudes $D$.
In the context of the last two paragraphs, 
it is worth mentioning that we can equivalently investigate
the wave-function $\psi$ written in its complex representation, 
$\psi = D \exp (i \varphi)$.

Thus, one concludes that the toy-model vacuum is filled with the 
set of vibrations which can be represented by standing homogeneous 
scalar plane waves in different inertial reference frames of 
the spacetime that is given 
by $a = 0$ assumed in the metric (\ref{sol}); it is clear that each 
of the investigated waves is characterized by different values of 
the wave parameters $\omega$ and $\bf k$ which, however, do not depend 
on the space and time coordinates for any single wave from the set. 
As is well-known from the properties of the (four-dimensional) Fourier 
transform, one can equivalently consider a set of standing, in general
inhomogeneous, scalar plane waves in a given reference frame of 
the spacetime $a = 0$; for such a reference system we can
choose any one of the inertial frames of the spacetime $a = 0$ --
obviously, in each of them (separately) the excitation corresponding 
to the cutoff mass $m_c \equiv E_c/c^2 = \hbar \omega_c/c^2$ remains
at rest with respect to the three-dimensional space $a = 0$, so
one has ${\bf p} = {\bf 0}$ for the cutoff excitation; see, 
for instance, sections 3 and 7.2 of this paper. Note also that in 
the case of an inhomogeneous wave, each of its wave parameters 
$\omega = \omega (t, {\bf r})$ and ${\bf k} = {\bf k} (t, {\bf r})$ 
can by definition depend on the space and time coordinates.

It is now interesting to note that the Lorentz transformation for 
the spacetime $a = 0$, or the four-vectors $(ct, {\bf r})$ and
$(\omega, c \, {\bf k}) = \pm (E, c \, {\bf p})/\hbar$ can easily be 
derived from first principles while considering the properties of the 
investigated field of the standing scalar plane waves filling the 
vacuum; see Refs.~\cite{Gleiser,Harter}. Similarly, one can obtain 
the Maxwell equations as well as the formula for the Lorentz 
electromagnetic force, both of which determine and describe the 
behaviour of the electromagnetic field that results from the 
deformation of the discussed standing scalar plane waves filling 
the vacuum; see Refs.~\cite{Cornil2,Cornil1} and references therein. 
In this section we will try to derive the Maxwell equations while 
investigating the properties of the phase function $\varphi = 
\varphi (t, {\bf r})$ of an inhomogeneous 
scalar plane wave; the proposed procedure closely follows 
that contained in section~6 of P.~Cornille's paper \cite{Cornil2} 
where one considers the differential form of the phase function 
$\varphi$, which in its general representation can be written as
\begin{equation}
d \varphi = \alpha \!\left( t, {\bf r} \right) \omega
\!\left( t, {\bf r} \right) dt - \alpha \!\left( t, {\bf r} \right)
{\bf k} \!\left( t, {\bf r} \right) \cdot d {\bf r} \ ,
\label{diff}
\end{equation}
with the function $\alpha = \alpha ( t, {\bf r} )$ being 
a dimensionless integrating factor; see also Ref.~\cite{Olariu} 
and references therein. The phase differential is a total one (so 
it can be integrated) if the two following conditions are fulfilled,
\begin{eqnarray}
\frac{\partial {\bf k}}{\partial t} + \nabla \omega = 
- a {\bf k} - {\bf b} \omega
\label{cond1} \\
\nabla \times {\bf k} + {\bf b} \times {\bf k} = 0 \ ,
\label{cond2}
\end{eqnarray}
where $\nabla \equiv \nabla_{\bf r}$ and
the quantities $a = a (t, {\bf r})$ and ${\bf b} = {\bf b} (t, {\bf r})$ 
are defined as $a \equiv \partial (\ln \alpha)/\partial t$ and 
${\bf b} \equiv \nabla (\ln \alpha)$, respectively (here the function 
$a$ should not be confused with the coordinate of the additional
spatial dimension). One also assumes that the phase factor, or Fourier
mode $\exp(i \varphi)$ is a solution of the homogeneous (source-free)
scalar wave equation, which results in the following formula:
\begin{equation}
\frac{1}{c^2} \frac{\partial \omega}{\partial t} + \nabla \cdot
{\bf k} = - \frac{a}{c^2} \omega - {\bf b} \cdot {\bf k} \ .
\label{cond3}
\end{equation}
After some algebra, one is able to obtain a set of the following 
equations regarding the electromagnetic four-vector potential 
$A^\mu \equiv (\Phi/c, {\bf A})$ where $\mu = 0, 2, 3, 4$ as well as 
$\Phi = \Phi (t, {\bf r})$ and ${\bf A} = {\bf A} (t, {\bf r})$,
\begin{eqnarray}
\nabla^2 \Phi - \frac{1}{c^2} \frac{\partial^2 \Phi}{\partial
t^2} \!\!\! &=& \!\!\! - \frac{\rho}{\epsilon_0}
\label{max1} \\
\nabla^2 {\bf A} - \frac{1}{c^2} \frac{\partial^2 {\bf A}}{\partial
t^2} \!\!\! &=& \!\!\! - \mu_0 {\bf J}
\label{max2} \\
\nabla^2 P - \frac{1}{c^2} \frac{\partial^2 P}{\partial
t^2} \!\!\! &=& \!\!\! 
\left( \frac{\mu_0}{\epsilon_0} \right)^{\! 1/2} 
\!\left( \frac{\partial \rho}{\partial t} + 
\nabla \cdot {\bf J} \right) \ ,
\label{max3}
\end{eqnarray}
with the modified Lorentz gauge
\begin{equation}
\frac{1}{c} \frac{\partial \Phi}{\partial t} + c \, \nabla \cdot
{\bf A} = - P 
\label{gauge}
\end{equation}
where the symbols $\rho = \rho (t, {\bf r})$ and ${\bf J} = {\bf J}
(t, {\bf r})$ stand for the density functions of the 
electric charge and the electric current, 
respectively, $\mu_0$ is the permeability of the vacuum, satisfying
the equality $c = (\mu_0 \epsilon_0)^{-1/2}$, and the 
quantity $P = P (t, {\bf r})$ denotes the so-called scalar 
polarization given by the relationship
\begin{equation}
P = \frac{c \hbar}{e} \!\left( \frac{a \omega}{c^2} +
{\bf b} \cdot {\bf k} \right) \ .
\label{polar}
\end{equation}
In turn, the self-consistency of the model being considered requires 
the following formula for the electric-field strength,
\begin{equation}
{\bf E} = \frac{\hbar}{e} (a {\bf k} + {\bf b} \omega) \ .
\label{elec}
\end{equation}
The electric-field and the magnetic-field strengths, given by 
the functions ${\bf E} = {\bf E} (t, {\bf r})$ and ${\bf B} = 
{\bf B} (t, {\bf r})$, are conventionally defined
with the use of the potentials $\Phi \equiv (\hbar/e) \omega$
and ${\bf A} \equiv (\hbar/e) {\bf k}$, so one has
${\bf E} = - \nabla \Phi - \partial {\bf A}/\partial t$ and ${\bf B} 
= \nabla \times {\bf A}$; see also the formulae (\ref{elec}),
(\ref{cond1}) and (\ref{cond2}). Additionally, both the electric-field 
and the magnetic-field strengths should fulfil the condition 
${\bf E} \cdot {\bf B} = (\hbar/e)^2 (a {\bf k} + {\bf b} \omega)
\cdot (\nabla \times {\bf k}) = 0$ where the latter equality results 
directly from Eq.~(\ref{cond2}). For the sake of completeness we give 
here the definitions of the functions $\rho = \rho (a, {\bf b},
\omega)$ and ${\bf J} = {\bf J} (a, {\bf b}, {\bf k})$, which 
have been used while deriving equations (\ref{max1})--(\ref{max3}),
\begin{eqnarray}
\frac{e}{\hbar \epsilon_0} \rho \!\! & \equiv & \!\! {\bf b} 
\cdot (\nabla \omega - {\bf b} \omega) + \nabla \cdot ({\bf b} \omega) 
- \frac{1}{c^2} \!\left[ \frac{\partial (a \omega)}{\partial t}
+ a \frac{\partial \omega}{\partial t} + a^2 \omega \right] 
\label{rho_def}  \\
\frac{e \mu_0}{\hbar} {\bf J} \!\! & \equiv & \!\!
{\bf b} (\nabla \cdot {\bf k} + {\bf b} \cdot {\bf k} ) + 
\nabla ({\bf b} \cdot {\bf k}) - \nabla \times ({\bf b} \times
{\bf k}) \nonumber \\  & \empty &  \hspace{1.823in}
- \frac{1}{c^2} \!\left[ \frac{\partial (a {\bf k})}{\partial t} +
a \frac{\partial {\bf k}}{\partial t} + a^2 {\bf k} \right] \ ;
\label{j_def}
\end{eqnarray}
one expects that the investigation of the above formulae and equation 
(\ref{max3}), carried out in the context of section 7.6 of this paper, 
could possibly shed new light on the origin and nature of electric 
charge. It is also interesting to remark that after performing some 
calculations concerning Eqs.~(\ref{cond1}), (\ref{cond3}) and 
assuming that the coefficients $a$ and $\bf b$ are constant with 
respect to both the space and time coordinates, one is able to 
conclude that the phase function $\varphi$ is a solution 
of the homogeneous scalar wave equation, 
$\nabla^2 \varphi - \partial^2 \varphi/\partial (c t)^2 = 0$,
with the additional relations ${\bf k} = - \exp (- a t - {\bf b} \cdot
{\bf r}) \, \nabla \varphi$ and $\omega = \exp (- a t - {\bf b} \cdot
{\bf r}) \, \partial \varphi/\partial t$ to be satisfied; note that
both the latter expressions remain then consistent with the formula 
(\ref{diff}).
 
Let us now assume that the value of the scalar polarization $P$ can
be different from zero only inside extended elementary particles; 
see Refs.~\cite{Gallop,Alicki,Cornil2} and references therein. 
The requirement that $P = 0$ is equivalent to the condition 
$v |{\bf b}| \cos ({\bf b}, {\bf n}) = -a$ provided that the 
dispersion relation $\omega = (c^2/v) |{\bf k}|$ holds, where the 
function $v = v (t, {\bf r})$ denotes the conventionally defined group 
velocity of the inhomogeneous wave(s) being considered, or its absolute 
value $|{\bf v}| \equiv v$; additionally, in the case when $v = c$ one
demands that $\cos ({\bf b}, {\bf n}) \neq \pm 1$ in order to prevent 
the electric field (\ref{elec}) from vanishing. On the other hand,
it is easy to prove that if we assume the conditions ${\bf n} \cdot
{\bf E} = 0$ and $\omega = c |{\bf k}|$ to be fulfilled, then we
obtain the equality $P = 0$, so indeed the value of the scalar
polarization $P$ is equal to zero, at least outside the sources
of the electromagnetic field; however, from the assumption that 
$v = c$ made implicitly in the above sentence it clearly results that 
the scalar plane waves representing the virtual excitations with 
non-zero masses $m \neq 0$ are not involved in the process under 
consideration, which is evidently consistent with the fact that 
we attempt to model here the phenomena of a purely electromagnetic 
nature. Then, for $P = 0$ equations (\ref{max1}) and (\ref{max2}) 
obviously remain the standard Maxwell (wave) equations for the 
electromagnetic potentials $\Phi$ and $\bf A$ 
with the field-sources characterized by the quantities $\rho$ and 
$\bf J$, respectively. In turn, the formulae (\ref{gauge}) and 
(\ref{max3}) become in such a case, or reduce to the standard Lorentz 
gauge and to the continuity equation for the four-vector $(c \rho, 
{\bf J})$ of the electric charge--current densities, respectively.

One then concludes that the electromagnetic phenomena occurring in 
the toy model can be the result of the deformation of the standing
scalar plane
waves filling the vacuum, which waves otherwise remain in a state of
equilibrium, with the phase function $\varphi$ being a total 
differential and with the phase factor $\exp (i \varphi)$ remaining 
a solution of the homogeneous scalar wave equation. It should be noted 
that the above-mentioned state of equilibrium is determined by the 
condition $\alpha (t, {\bf r}) = {\it const} > 0$ assumed in the 
differential form (\ref{diff}) of the phase function $\varphi$; 
in such a case one has $a = 0$ and ${\bf b} = {\bf 0}$, which implies 
that the values of the electric-field strength $\bf E$ defined 
by expression (\ref{elec}) as well as 
the magnetic-field strength $\bf B$ are both equal to zero. 
Any perturbation of the equilibrium state, which originates in the 
sources characterized by the densities of the electric charge $\rho$ 
and the electric current $\bf J$, induces -- in the vacuum -- a new set
of standing scalar plane waves associated with a progressive 
transverse (vector) plane wave, and the latter wave 
is actually an electromagnetic one; see equations (\ref{max1}) and
(\ref{max2}) as well as the definitions of the quantities $\Phi$,
$\bf A$, $\bf E$ and $\bf B$, contained in the sentence appearing 
directly below the formula (\ref{elec}).

In the context of this section one can add as well that
the topology of the toy-model manifold requires a massless particle
to have the same, constant velocity in any inertial (Lorentz) reference
frame of the spacetime $a = 0$; see Ref.~\cite{wt}. From the toy model 
presented in this paper it also results clearly that the phenomenon
of the very high, but finite value of the speed of light is closely 
related to the existence of the Planck-frequency cutoff imposed on 
the field of the standing scalar plane waves which fill the 
vacuum; namely, the velocity of the energy exchange -- occurring at 
the microscopic level, between the scalar wave-field and the transverse
electromagnetic waves, in the vacuum -- depends on the sizes
of the basic ``grains'' of the spacetime, which sizes are of the order 
of the Planck time $T_{\it Pl}$ and of the Planck length $L_{\it Pl}$.

\bigskip
\bigskip
\noindent {\bf 8.\ Dimensionless Einstein equations}
\bigskip

According to the toy model, one has two natural units imposed on
the spacetime:\ these are the cutoff units for time and space, which
are of the order of the Planck time $T_{\it Pl}$ and of the Planck 
length $L_{\it Pl}$, respectively. If we expect the quantum theory of
gravity to exist at all, then we can suppose that the quantum effects 
appear at the Planck scale of time and length. Therefore, it seems
reasonable to rescale the quantities $x^\mu$ and to introduce new
(dimensionless) spacetime coordinates ${\widetilde x}^\mu \equiv 
x^\mu/L_c$ for $\mu = 0, \ldots, 4$ where $x^0 \equiv ct$. Thus, the 
variables ${\widetilde x}^\mu$ for $\mu = 0, \ldots, 4$ should be 
of the order of unity for effects proper to the quantum gravity.
After performing the above transformation of the coordinates, the 
metric tensor takes
the form ${\widetilde g}_{\mu\nu} ( {\widetilde x}^\mu) =
g_{\mu\nu} ( x^\mu ) \equiv g_{\mu\nu} 
( L_c \, {\widetilde x}^\mu )$ for $\mu, \nu = 0, \ldots, 4$. 
We also define $\gamma^\alpha_{\mu\nu} ({\widetilde g}_{\mu\nu}, 
{\widetilde x}^\mu) \equiv L_c \Gamma^\alpha_{\mu\nu}
(g_{\mu\nu}, x^\mu)$ and $r_{\mu\nu} ({\widetilde g}_{\mu\nu},
{\widetilde x}^\mu) \equiv L^2_c R_{\mu\nu} (g_{\mu\nu},
x^\mu)$ as well as $t_{\mu\nu} ( {\widetilde g}_{\mu\nu},
{\widetilde x}^\mu) \equiv 8 \pi G T_{\mu\nu} (g_{\mu\nu}, 
x^\mu)/(c^4 |\lambda|)$ and $r \equiv r_{\mu\nu} 
{\widetilde g}^{\mu\nu}$, all for
$\alpha, \mu, \nu = 0, \ldots, 4$. Substituting the above
quantities into Eqs.~(\ref{1}), we then obtain easily the
dimensionless Einstein field equations devoid of any physical 
constants,
\begin{equation}
r_{\mu\nu} - \frac{1}{2} {\widetilde g}_{\mu\nu} r 
- 4 \pi^2 {\widetilde g}_{\mu\nu} = - 4
\pi^2 t_{\mu\nu} \ .
\label{dless}
\end{equation}
It is clear that the above set of equations possesses solutions which
contain no dimensional quantities, but only pure numbers and
dimensionless coordinates. One has to rescale such solutions 
-- with the use of the transformation given above which incorporates 
the physical constants $\hbar$, $c$ and $G$ -- in order to obtain the
corresponding solutions of Eqs.~(\ref{1}). Thus, in such an approach
the set of equations (\ref{dless}) can be regarded as being, in a 
sense, primary with respect to the set of equations (\ref{1}).

The question then arises as to whether we can in the above-defined 
transformation use a set of the fundamental constants 
other than $\hbar$, $c$ and $G$ (or, equivalently, other than the 
quantities $T_c$, $L_c$ and $E_c$), in order to obtain from 
Eqs.~(\ref{dless}) a set of equations describing interation fields 
other than gravity, e.g., the electroweak and/or the strong ones. 
The tensor $g_{\mu\nu}$ is expected to be there in the form of a 
tensor of ``potentials'', with its components characterizing the 
above-mentioned interactions (such as the four-vector potential $A^\mu$ 
in the case of the electromagnetic forces); those potentials would be 
obtained from an appropriate contraction of the tensor $g_{\mu\nu}$. 
Surely, such solutions of Eqs.~(\ref{dless}) could be subject to some 
quantization-like procedures. We should recall here that the 
considerations quite similar to those described above lead one to the
formulation of bi-metric, or bi-scale theories of gravitation and
elementary particles (e.g., concerning also the strong interactions 
between particles); such models have been proposed, among others, 
by N.~Rosen and by A.~Salam; see, for instance, Ref.~\cite{Recami} 
and references therein.

\bigskip
\bigskip
\noindent {\bf 9.\ Dimensionless Lagrangian and the coupling constant}
\bigskip

Let us now write an action $S$ whose extremization with respect
to the metric tensor ${\widetilde g}^{\mu\nu}$ leads to the
obtaining of Eqs.~(\ref{dless}),
\begin{equation}
S \equiv S_m + S_g = \int \!\left( {\cal L}_m + 
{\cal L}_g \right) d^5 {\widetilde x} \ ,
\label{action}
\end{equation}
where $S_m$ and $S_g$ are the matter and the Einstein--Hilbert
gravitational actions obtained by the integration (over all the 
spacetime coordinates) of the matter and the gravitational 
Lagrangians, ${\cal L}_m$ and ${\cal L}_g$, respectively;
note that the matter Lagrangian ${\cal L}_m$ contains also the
contribution from the zero-point fields of the vacuum. We have
\begin{equation}
{\cal L}_g = \frac{\left\vert {\rm det} \!\left( 
{\widetilde g}_{\mu\nu} \right) \right\vert^{1/2}}{8 \pi^2} 
\!\left( r + 8 \pi^2 \right) \ ,
\label{Lagr}
\end{equation}
which expression represents the gravitational Lagrangian with the
dimensionless coupling constant $4 \pi$ instead of the ``usual''
gravitational coupling constant $(32 \pi G/c^3)^{1/2}$;
see also Refs.~\cite{Feynman,Strom}. Of course, the above na\"{\i}ve
renormalization of the coupling constant is a consequence of the
fact that the energy density of the ground state (i.e., of the
vacuum) takes a finite and well-defined value in our model. One 
should then add that in the sense usual for the quantum theory, a 
renormalization procedure in the toy model consists in operating 
(e.g., subtracting from each other) on Planck-scale order, finite 
quantities rather than on arbitrarily infinite ones.

The presence of the Newtonian constant $G$ of the 
dimensionality of a negative power of mass in the ``usual''
gravitational Lagrangian is regarded as one of the main reasons
implying the non-renormalizability of the quantum gravity; see, for
instance, Refs.~\cite{Goroff,DesY}. For the gravitational Lagrangian
containing the constant $G$, each order of the perturbation theory
(PT) will give rise to a new counter-term with a new dimensionality.
Thus, if only one order of the PT is non-renormalizable, like  
the two-loop divergences in Ref.~\cite{Goroff}, then such is the 
whole perturbation theory, too. However, when the gravitational 
Lagrangian is dimensionless, then one can at least hope that the 
non-renormalizable term(s) arising in some order(s) of the PT will 
be cancelled out by the other non-renormalizable term(s) appearing
in another order(s) of the perturbation theory. Thus, we should
consider whether the formulation of the gravitational Lagrangian
${\cal L}_g$ in the dimensionless form given by expression
(\ref{Lagr}) can allow one to reach any new conclusions suggesting 
how to renormalize the quantized pure Einstein gravity
\cite{Goroff} as well as the quantized Einstein--Yang--Mills 
\cite{DesY,Kall}, Einstein--Maxwell \cite{DesE} and 
Dirac--Einstein \cite{DesD} systems. On the other hand, we already 
know that the introduction of additional spatial dimension(s) into
the spacetime manifold leads one to the obtaining of finite results 
concerning, at least, the lowest-order ``divergent'' graphs in the 
quantum field theory; see Refs.~\cite{Boll,Ashm,Cict,Hooft,Capp,Engl}. 
It is clear that many efforts concerning both the above subjects are 
needed within the toy model under consideration to obtain any 
satisfactory and binding results.

\newpage

\centerline{\epsfxsize=2.85 true in \epsfbox{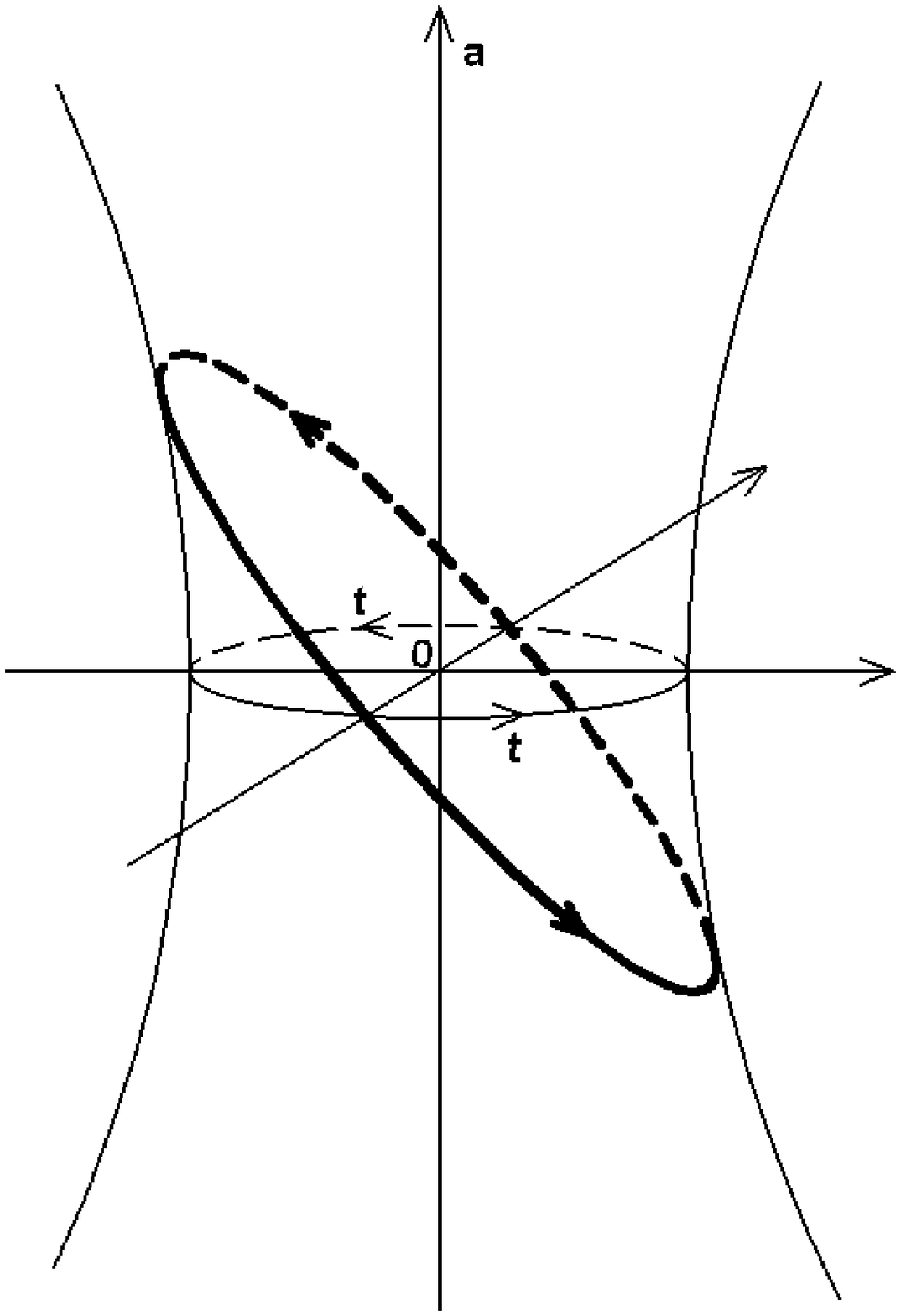}}

\smallskip

\noindent {\small {\bf Figure 1.}\ The toy model predicts that 
the Universe before the Big Bang was of the form of the anti-de Sitter 
spacetime ${\bf S}^1 \times {\bf R}^1 \times {\bf R}^3$ filled with 
the radiation quanta, each of which possessed the energy $E$ equal 
to the cutoff energy $E_c$. The geodesic lines of the radiation
particles, or quanta (the thick line in the figure represents one 
of them) were global closed null curves, with the periods of 
(a single) oscillation all equal to the cutoff time 
$T_c = 2 \pi/(c |\lambda|^{1/2})$. 
Thus, the radiation quanta retraced their own life histories after 
each lapse of the period $T_c$ of the coordinate time $t$. At the 
moment of the Big Bang, the phase transition ${\bf S}^1 \to {\bf R}^1$ 
of the coordinate time occurred, which gave the beginning to the 
expansion of the flat three-dimensional space ${\bf R}^3$ due to the 
release of the radiation into that space ${\bf R}^3$. Note that the toy
model assumes that no particle can possess the energy $E$ greater 
than the cutoff energy $E_c = h/T_c$. After performing the calculation
of the value of the cosmological constant $\lambda$ in this paper, it
turns out that the cutoff energy $E_c$ is of the order of the Planck
energy $E_{\it Pl}$, as one might have expected and as it was assumed
{\it a priori} in Ref.~\cite{wt}. {{\bf{\em Note.}}\ For reasons of 
clarity, only the anti-de Sitter two-dimensional spacetime ${\bf S}^1
\times {\bf R}^1$ is shown, so the figure does not incorporate 
the flat space ${\bf R}^3$. The coordinate of the additional spatial 
dimension is denoted by $a$; see Ref.~\cite{wt} for more details.}

\end{document}